\documentclass{article}

\usepackage[backend=biber,style=numeric,sorting=none,citestyle=numeric-comp]{biblatex}
\renewbibmacro{in:}{}
\usepackage{setspace}
\usepackage[english]{babel}
\usepackage[utf8]{inputenc}
\usepackage{amsmath}
\usepackage{amsfonts}
\usepackage{amssymb}
\usepackage{amsthm}
\usepackage{cancel}
\usepackage{slashed}
\usepackage{graphics}
\usepackage{graphicx}
\usepackage{fancyhdr}
\usepackage{multirow}
\usepackage{float}
\usepackage{breqn}
\usepackage{pdfpages}
\usepackage{lipsum}

\usepackage{hyperref}
\usepackage{fancyhdr}
\usepackage{bm}

\usepackage[left=3cm,top=2.25cm,bottom=2.75cm,right=2.5cm,includehead]{geometry}
\usepackage[titletoc]{appendix}
\usepackage{youngtab}
\usepackage{dsfont}
\usepackage{tikz}
\usetikzlibrary{shapes,arrows}
\usepackage{bbm}
\usepackage{verbatim}
\usepackage{indentfirst}
\usepackage{xspace}

\usepackage{physics} 
\usepackage{subcaption}

\newcommand{\CMM}{$\mathbb{C}$MM\xspace}
\newcommand{\CLM}{$\mathbb{C}$LM\xspace}
\newcommand{\CMMs}{$\mathbb{C}$MMs\xspace}
\newcommand{\CLMs}{$\mathbb{C}$LMs\xspace}

\hypersetup{
    colorlinks=true,
    linkcolor=blue,
    urlcolor=cyan,
    citecolor=cyan,}

\title{Pseudospectra of complex momentum modes}
\author{David García Fariña$^{1,2}$, Karl Landsteiner$^1$, \\  Pau G. Romeu$^{1,2}$ and Pablo Saura-Bastida${}^{1,3}$}

\newtheorem{theorem}{Thm.}[section]

\newtheorem{definition}{Def.}[section]

\begin{document}

\maketitle\thispagestyle{empty}
{\begin{center}${}^1$Instituto de F\'isica Te\'orica UAM/CSIC, c/Nicol\'as Cabrera 13-15, Universidad Aut\'onoma de Madrid, Cantoblanco, 28049 Madrid, Spain\\
${}^2$Departamento de F\'isica Te\'orica, Universidad Aut{\'o}noma de Madrid, Campus de Cantoblanco, 28049 Madrid, Spain\\
${}^3$ Departamento de Autom\'atica, Ingenier\'ia El\'ectrica y Tecnolog\'ia Electr\'onica, Universidad Polit\'ecnica de Cartagena, Calle Dr. Fleming, S/N, 30202 Cartagena, Murcia 
\end{center}}

\begin{abstract}
    \par
  We initiate the study of stability and pseudospectra of complex momentum modes of \mbox{asymptotically} anti-de Sitter black holes. Similar to quasinormal modes, these can be defined as the poles of the holographic Green’s function, albeit for real frequency and complex momentum. Their pseudospectra are in stark contrast to the pseudospectra of quasinormal modes of AdS black holes. Contrary to the case of quasinormal mode pseudospectra, the resolvent is well-defined, and the numerical approximation shows fast convergence. At zero frequency, complex momentum modes are stable normal modes of a Hermitian operator. Even for large frequencies, they show only comparatively mild spectral \mbox{instability}. We also find that local potential perturbations cannot destabilize the lowest complex momentum mode.

\end{abstract}
\vfill
{\tt
\href{mailto:david.garciafarinna@estudiante.uam.es}{david.garciafarinna@estudiante.uam.es}\\[-5pt] 
\href{mailto:pau.garcia@uam.es}{pau.garcia@uam.es}\\[-5pt] 
\href{mailto:karl.landsteiner@csic.es}{karl.landsteiner@csic.es}\\[-5pt] 
\href{mailto:pablo.saura@upct.es}{pablo.saura@upct.es}
}
{\hfill IFT-UAM/CSIC-24-100 }

\newpage
\tableofcontents

\newpage

\section{Introduction}\label{sect:Introduction}

Eigenvalue problems for linear operators are fundamental to all of physics, from classical mechanics to quantum mechanics and general relativity. The eigenvalues are of course determined by the particular boundary conditions. Closed systems lead to the emblematic Sturm-Liouville type problems for self-adjoint differential operators.
Open systems on the other hand are characterized by non-self-adjoint and potentially non-normal operators. Non-normality implies that under a small perturbation to the operator, the eigenvalues might suffer a displacement much larger than the size of the perturbation. This phenomenon, labelled spectral instability, is a fundamental feature of non-normal operators. For normal operators, such as the ones appearing in conservative systems, the spectral theorem ensures that the eigenvalues cannot be displaced more than the size of the perturbation, and thus the spectrum is stable.

At a practical level, spectral instability has significant implications. It tells us that if we use some non-normal operator to model a non-conservative system and assume a small error in its construction (e.g. disregarding some interactions to simplify computations), we cannot ensure that the spectrum of the operator should match the actual spectrum of the real system. In short, the presence of spectral instability warns us that to make predictions to a given accuracy, we will need to construct a model with a much higher one. For that reason, when dealing with non-conservative systems, one should complement the standard eigenvalue analysis with a discussion on the stability of the eigenvalues, which typically is approached through the study of the pseudospectrum \cite{Trefethen:2005, Sjostrand:2019,davies:2007}.

In the context of gravitational physics, non-self-adjoint operators generically appear when studying quasinormal modes (QNMs) of black holes \cite{Nollert:1999ji, Kokkotas:1999bd, Berti:2009kk}. The non-conservative nature of black hole geometries, derived from the existence of an event horizon which acts as a perfectly absorbing membrane, results in the time evolution of linearized fluctuations being governed by a non-self-adjoint differential operator, whose eigenvalues are the quasinormal frequencies (QNFs). In asymptotically flat spacetimes, fluctuations are also outgoing at asymptotic infinity further contributing to the non-self-adjointness.

The study of the spectral stability of quasinormal frequencies using a pseudospectrum analysis has been recently pioneered in \cite{Jaramillo:2020tuu}. Further aspects of the pseudospectrum of quasinormal modes of asymptotically flat black holes have been investigated in \cite{Destounis:2021lum, Cheung:2021bol, Berti:2022xfj, Konoplya:2022pbc, alsheikh:tel-04116011, Destounis:2023ruj} (see \cite{Boyanov:2022ark} also for horizonless compact objects and \cite{Cao:2024oud} for quantum corrected black holes), as well as in de Sitter (dS) \cite{Sarkar:2023rhp, Destounis:2023nmb} and anti-de Sitter (AdS) \cite{Arean:2023ejh, Cownden:2023dam, Boyanov:2023qqf} geometries. All these studies found the spectrum of quasinormal frequencies to be unstable due to the non-conservative nature of the geometry. Further studies have also covered the implications of this spectral instability on the strong cosmic censorship \cite{Courty:2023rxk} and on the gravitational waves emitted in black hole mergers \cite{Jaramillo:2021tmt, Jaramillo:2022kuv}.

In the case of asymptotically AdS spacetimes, the observation of spectral instability is of particular interest in view of gauge/gravity duality \cite{Maldacena:1997re, Aharony:1999ti, Ammon:2015wua, zaanen2015holographic, Hartnoll:2018xxg} which postulates that black hole geometries are dual to thermal states in a strongly coupled quantum field theory \cite{Witten:1998zw}. In this context, black hole quasinormal frequencies are interpreted as poles of retarded propagators of the thermal field 
theory \cite{Horowitz:1999jd, Birmingham:2001pj, Kovtun:2005ev}. Consequently, the study of quasinormal frequencies in AdS black holes has led to important insights into hydrodynamics and transport theory in the relativistic regime \cite{Policastro:2001yc, Baier:2007ix, Bhattacharyya:2007vjd}, with some remarkable results being the extremely low specific shear viscosity of holographic models of the quark-gluon plasma \cite{Kovtun:2004de}, phase transitions towards superconducting states \cite{Gubser:2008px, Hartnoll:2008vx, Amado:2009ts, Herzog:2009ci} and strongly coupled quantum critical phases \cite{Herzog:2007ij, Cubrovic:2009ye, Iqbal:2011ae}. 
To better understand the physical meaning of this spectral instability is therefore clearly of utmost importance for gauge/gravity duality.

Quasinormal frequencies determine the time evolution of an open system that is slightly perturbed away from its ground state. Figuratively one hits the system with a ``hammer'' and observes the damped oscillations that return the system to its ground state. This is however not the only way of probing an open or dissipative system. Another way is to couple it locally to an oscillating source (``antenna'') and observe the wavelength and absorption length of the forced oscillation. This type of response is also determined by the poles of the retarded Green's function, this time however one fixes the frequency to be real and searches for poles in the complexified momentum plane. In gauge/gravity duality this means that one solves the same system of equations as for the quasinormal modes but considers now the frequency as given and solves for the complex eigenvalues of the momentum \cite{Amado:2007pv}. These complex momentum modes are the AdS analogues of the complex angular momentum modes in asymptotically flat spacetimes \cite{Andersson:1994rk}. We denote complex momentum modes by \CMMs and their corresponding eigenvalues, the complex (linear) momenta by \CLMs.

In this paper, we therefore depart from the study of the stability of QNFs to instead consider the stability of \CLMs of planar AdS black holes. While the former are eigenvalues of the generator of time translations, the latter correspond to eigenvalues of the generator of spatial translations along some direction parallel to the brane. QNFs and \CLMs are dual to poles of the retarded Green's function at fixed momentum (relaxation times) and fixed frequency (absorption lengths), respectively. \CLMs also play an essential role in studying causality of the boundary field theory \cite{Landsteiner:2012gn, Gavassino:2023mad}. At zero frequency, they are dual to the glueball masses of a dimensionally compactified toy model for QCD \cite{Witten:1998zw, Csaki:1998qr, deMelloKoch:1998vqw, Brower:2000rp, Bak:2007fk}. \CMMs have also appeared in the holographic context in \cite{Sonner:2017jcf,Novak:2018pnv,Heller:2020uuy,Janik:2021jbq}. It is also worth noting that in the context of asymptotically flat spacetimes \cite{Torres:2023nqg,Rosato:2024arw, Oshita:2024fzf} studied the stability of Regge poles (complex angular momenta).

We choose to study the stability of complex momenta for two main reasons
\begin{itemize}
    \item \textbf{Complex momenta offer a new window to probe the spectral stability of the theory}. Although both \CLMs and QNFs are poles of the retarded Green's function, they appear in qualitatively very different settings, and thus, their spectral properties need not be the same. Hence, we could find a theory whose QNFs are stable while the \CLMs are not and vice versa. 
    
    \item \textbf{The pseudospectra of complex momenta computed at real frequencies is convergent}. When computing the pseudospectra of QNFs, one finds that it does not converge for the overtones \cite{Boyanov:2023qqf}. This issue seems to be related to the definition of size for the perturbations and, as we shall see in greater detail, hints at the need to introduce a cutoff to how localized the perturbations can be. Notably, this prevents us from making quantitative statements about the stability, although it is believed that the qualitative picture is correct. On the other hand, working at fixed real frequency solves these issues and thus allows us to reach quantitative conclusions about the full pseudospectrum.
\end{itemize}

We compute the \CLM pseudospectrum for a real scalar field in a Schwarzschild AdS$_{4+1}$ black brane. Although this setup lacks some interesting phenomenology associated with the existence of hydrodynamic modes, we favour its simplicity and lack of gauge symmetries, which would introduce some extra subtleties. We leave a more complete study for fields with gauge symmetries to a follow-up work.

The paper is structured as follows. In section \ref{sect:ReviewPseudo} we offer a short review of pseudospectra in the context of stability analysis. Here, we follow closely \cite{Arean:2023ejh} and draw heavily from \cite{Trefethen:2005}. 

In section \ref{sect:ConvergencePseudoQNFs} we discuss the lack of convergence of the pseudospectrum of QNFs. Our presentation is not in-depth; instead, we try to offer a more qualitative picture. We refer the interested reader looking for a more detailed discussion to \cite{Boyanov:2023qqf}. 

In section \ref{sect:Complex momentum modes}, we define \CMMs and showcase, from the point of view of the dual field theory, in what situations they appear. We discuss their differences with respect to QNFs and conclude by arguing some of the stability properties complex momenta should have. In particular, in the zero frequency limit, we argue that as \CLMs are dual to the glueball spectrum of the effective theory for the lowest Kaluza-Klein modes of the compactification on a thermal circle, they should be stable.

In section \ref{sect:Holographic model}, we introduce the holographic model and construct a well-motivated notion of size for the \CMMs. We follow the usual prescription introduced in \cite{Jaramillo:2020tuu, Gasperin:2021kfv} and more generally argued for in \cite{Trefethen:2005}, and use the energy as a guide to construct a norm in subsection \ref{subsect:EnergyNorm}. In subsection \ref{subsect:ConstructionL}, we define the relevant differential operator whose eigenvalues are the \CLMs, and construct its adjoint. We pay special attention to how our norm manages to reproduce the expected stability properties discussed in section \ref{sect:Complex momentum modes}. 

In section \ref{sect:Numerical method} we discuss the numerical implementation based on pseudospectral methods and introduce the selective pseudospectrum used to test the stability under random local potential perturbations. We also construct some specific potentials to test the nature of the (in)stability.

Section \ref{sect:Results} is the core of the paper and contains the results of our pseudospectrum computations. We find that at non-zero frequency, the spectrum becomes rapidly unstable. Remarkably, we observe that the effect of local potential perturbations seems to be very mild. 

In section \ref{sect:Conclusions}, we summarize our findings and present our conclusions.

In appendix \ref{app:cmms} we present the numerical values of the complex momenta. Relevant figures are collected in appendix \ref{app:PseudoPlots}. A comparison between the stability properties of \CLMs and QNFs can be found in appendix \ref{app:Comparisons}.

\section{Pseudospectra and Stability}\label{sect:ReviewPseudo}

In this section, we summarize some important definitions and results concerning pseudospectra and condition numbers and their relevance for the study of the spectral stability of non-normal operators. We follow closely the exposition of \cite{Arean:2023ejh} and refer the interested reader to \cite{Trefethen:2005} for further details.

Given a closed linear operator $\mathcal{L}$ acting on a Hilbert space $\mathcal{H}$ with domain $\mathcal{D}(\mathcal{L})$, its spectrum $\sigma(\mathcal{L})$ is defined as the set of points $\lambda$ in the complex plane where the resolvent $\mathcal{R}(\mathcal{L};\lambda)=(\mathcal{L}-\lambda)^{-1}$ is not defined. An eigenvalue $\lambda\in\sigma(\mathcal{L})$ is defined as a solution to the eigenvalue equation
\begin{equation}
    (\mathcal{L}-\lambda)u_\lambda=0 \,,
\end{equation}
where $u_\lambda\in\mathcal{D}(\mathcal{L})$ is the corresponding eigenvector. 

An important property of eigenvalues is that for self-adjoint adjoint operators (or, in more physical terms, conservative systems), the spectral theorem ensures that if we perturb the system with a bounded operator of size $\varepsilon$ the eigenvalues of the perturbed operator cannot suffer a displacement greater than $\varepsilon$ \cite{kato2013perturbation, Courant:1989}. This property holds in general for any normal operator $A$ satisfying $\left[A,A^\dagger\right]=0$, with $A^\dagger$ the adjoint of $A$. 

However, in non-conservative systems normality of the relevant operators is not guaranteed and small perturbations could potentially alter the spectrum in a significant manner. For that reason, one concludes~\cite{Trefethen:2005} that in non-conservative systems eigenvalue analysis alone is insufficient as the spectrum might be unstable.

To characterize the stability of eigenvalues one introduces the notion of $\varepsilon$-pseudospectrum, which can be defined in three mathematically equivalent ways \cite{Trefethen:2005}:
\begin{definition}[Resolvent norm approach]
    \label{def:Pseudo Definition 1}
    Given a closed linear operator $\mathcal{L}$ acting on a Hilbert space $\mathcal{H}$ with domain $\mathcal{D}(\mathcal{L})$, and $\varepsilon>0$, the $\varepsilon$-pseudospectrum $\sigma_\varepsilon(\mathcal{L})$ is
    \begin{equation}\label{eq:Pseudo Definition 1}
        \sigma_\varepsilon(\mathcal{L})=\{z\in \mathbb{C} : \lVert \mathcal{R}(z;\mathcal{L}) \rVert>1/\varepsilon \}\,,
    \end{equation}
    with the convention $\lVert \mathcal{R}(z;\mathcal{L}) \rVert=\infty$ for $z\in\sigma(\mathcal{L})$.
\end{definition}

\begin{definition}[Perturbative approach]
    \label{def:Pseudo Definition 2}
    Given a closed linear operator $\mathcal{L}$ acting on a Hilbert space $\mathcal{H}$ with domain $\mathcal{D}(\mathcal{L})$, and $\varepsilon>0$, the $\varepsilon$-pseudospectrum $\sigma_\varepsilon(\mathcal{L})$ is
    \begin{equation}\label{eq:Pseudo Definition 2}
        \sigma_\varepsilon(\mathcal{L})=\{z\in \mathbb{C}, \exists V,\rVert V \lVert<\varepsilon : z\in\sigma(\mathcal{L}+V) \}\,.
    \end{equation}
\end{definition}

\begin{definition}[Pseudoeigenvalue approach]
    \label{def:Pseudo Definition 3}
    Given a closed linear operator $\mathcal{L}$ acting on a Hilbert space $\mathcal{H}$ with domain $\mathcal{D}(\mathcal{L})$, and $\varepsilon>0$, the $\varepsilon$-pseudospectrum $\sigma_\varepsilon(\mathcal{L})$ is
    \begin{equation}\label{eq:Pseudo Definition 3}
        \sigma_\varepsilon(\mathcal{L})=%
        \{z\in \mathbb{C},\exists u^\varepsilon \in \mathcal{D}(\mathcal{L}) :\; \rVert (\mathcal{L}-z ) u^\varepsilon\lVert<\varepsilon\lVert u^\varepsilon \rVert\}\,,
    \end{equation}
    where $u^\varepsilon$ is a $\varepsilon$-pseudoeigenvector with $\varepsilon$-pseudoeigenvalue $z$.
\end{definition}

Note that, contrary to the spectrum, the pseudospectrum depends on the operator norm
\begin{equation}
    \rVert V \lVert=\max_{u\in\mathcal{H}}\frac{\rVert Vu \lVert}{\rVert u \lVert}\,,
\end{equation}
as, in order to quantify stability, it needs a notion of what constitutes a small perturbation.

Remarkably, definition \ref{def:Pseudo Definition 2} corresponds to the physical intuition we were seeking: the $\varepsilon$-pseudospectrum constitutes the maximal region containing all possible displacements of the eigenvalues under perturbations of size $\varepsilon$. It is then quite natural to represent the pseudospectrum as a contour map indicating the boundaries of these regions for multiple values of $\varepsilon$. However, computationally definition \ref{def:Pseudo Definition 2} is very inefficient as one should compute the spectra for all possible bounded perturbations. For that reason, when studying general instability properties one instead uses definition \ref{def:Pseudo Definition 1} which, despite lacking such a clear physical interpretation, is much more manageable to compute in a finite-dimensional setting such as the ones arising when employing numerical approximations to the operators. When constructing a selective pseudospectrum where we only analyze the stability under a certain restricted type of perturbations (e.g. local potentials), we will still use definition \ref{def:Pseudo Definition 2} with a finite set of randomly generated perturbations. 

Another useful tool for studying the stability of eigenvalues is the set of condition numbers $\{\kappa_i\}$ defined as
\begin{equation}
        \kappa_i=\frac{\norm{v_i}\norm{u_i}}{|\expval{v_i,u_i}|}\,,
\end{equation}
with $\expval{\cdot,\cdot}$ the inner product associated with the norm $\norm{\cdot}$, $u_i$ the right-eigenvector satisfying $\mathcal{L} u_i=\lambda_i u_i$ and $v_i$ the left-eigenvector satisfying $\mathcal{L}^\dagger v_i=\bar{\lambda}_i v_i$.\footnote{We denote the complex conjugate with a bar and complex conjugate transpose with an asterisk.} Remarkably, condition numbers manage to quantify the effect of perturbations of size $\varepsilon$ through only the knowledge of the orthogonality between the eigenvectors of the unperturbed operator and its adjoint. Explicitly, for a bounded perturbation $V$ of size $\varepsilon$ we have:
\begin{equation}
    |\lambda_i(\varepsilon)-\lambda_i|\leq \varepsilon  \kappa_i \,,
\end{equation}
where $\{\lambda_i(\varepsilon)\}$ are the eigenvalues of the perturbed operator $\mathcal{L}(\varepsilon)=\mathcal{L}+V$ and $||V||=\varepsilon$. For normal operators all eigenvalues are stable and have condition number 1.
    
As in this work, we limit ourselves to studying pseudospectra of matrices arising from the discretization of differential operators, we collect below a relevant theorem specialized to matrices, which allows to efficiently compute the pseudospectrum numerically. Again further details can be found on \cite{Trefethen:2005}.
   
\begin{theorem}
    \label{th:psuedospectrum in different norms}
    Given the $\ell^2_N$-inner product $\expval{\cdot,\cdot}_2$ 
    \begin{equation}
        \expval{v,u}_2=\bar{v}^i u^i \,,
    \end{equation}
    and generic $G$-inner product $\expval{\cdot,\cdot}_G$  such that:
    \begin{equation}
        \label{eq:G-norm in th:psuedospectrum in different norms }  
        \expval{v,u}_G= G_{ij} \bar{v}^i u^j\,,
    \end{equation}
    with $G=F^*F$ a symmetric positive definite $N\times N$ matrix.
    \begin{itemize}
        \item The $\varepsilon$-pseudospectrum of a matrix $M$ in the $G$-norm $\sigma_\varepsilon^G(M)$ satisfies
        \begin{equation}
        \label{eq:pseudo relation in th:psuedospectrum in different norms }      \sigma_\varepsilon^G(M)=\sigma_\varepsilon^{\ell^2_N}\left(FMF^{-1}\right)\,,
        \end{equation}
        with $\sigma_\varepsilon^{\ell^2_N}(M)$ the pseudospectrum in the $\ell^2_N$-norm.
        
        \item The condition number of the eigenvalue $\lambda_i$ of a matrix $M$ in the G-norm $\kappa_i^G$ satisfies:
        \begin{equation}
          \kappa_i^G=\frac{\norm{\tilde{v}_i}_2\norm{\tilde{u}_i}_2}{|\expval{\tilde{v}_i,\tilde{u}_i}_2|}\,,
        \end{equation}
        where $\tilde{u}_i$ and $\tilde{v}_i$ fulfill:
        \begin{equation}
            FMF^{-1}\tilde{u}_i=\lambda_i \tilde{u}_i\,,\qquad \left(FMF^{-1}\right)^*\tilde{v}_i=\bar{\lambda}_i \tilde{v}_i\,.
        \end{equation}
    \end{itemize}
    
\end{theorem}

Lastly, we note that throughout this section we have assumed $\varepsilon$ to be small. Formally this corresponds to saying that $\varepsilon$ is much smaller than the minimum distance $d_{\min}$ between disconnected regions of the spectrum, i.e., $\varepsilon/d_{\min}\ll 1$.

\section{Convergence of Pseudospectrum of Quasinormal Modes}\label{sect:ConvergencePseudoQNFs}

In this section we offer a short qualitative discussion on the lack of convergence of the pseudospectrum of quasinormal modes reported in \cite{Boyanov:2023qqf}. 

As pointed out in \cite{Warnick:2013hba}, when defining QNMs in regular coordinates one needs to demand analyticity on the event horizon. Otherwise, outgoing and ingoing modes would be indistinguishable and any frequency would then be a quasinormal frequency. This implies that if we define QNFs as eigenvalues of a non-normal operator acting on some Hilbert space, we should ensure that the norm of the said Hilbert space does indeed eliminate all non-analytic modes, i.e. that their norm is infinite.

To construct the pseudospectrum, we would like to have a physically motivated norm. For QNMs, as argued in \cite{Jaramillo:2020tuu,Gasperin:2021kfv}, such norm is given by their energy which typically contains at most two derivatives. However we should check whether such norm does discard all non-analytic modes.
Consider the non-analytic function $\psi(x)=x^{3/2}$. Then for a norm  containing only two derivatives (such as the energy)
\begin{equation}
    \norm{\psi}^2=\int_0^1 dx\,\left[ a\partial_x\psi\partial_x\bar{\psi}+b\psi\bar{\psi}\right]=\int dx\,\left[ a\frac{9}{4}\left(x^{1/2}\right)^2+b\left(x^{3/2}\right)^2\right]=\frac{9}{8}a+\frac{1}{4}b<\infty\,.
\end{equation}
Since $\norm{\psi}<\infty$ our construction  fails to remove this non-analytic mode. On the other hand, if we consider a norm with four derivatives such as
\begin{equation}
    \norm{\psi}^2=\int_0^1 dx\,\left[ a\partial_x\psi\partial_x\bar{\psi}+b\psi\bar{\psi}+c\partial_x^2\psi\partial_x^2\bar{\psi}\right]=\int dx\,\left[ a\frac{9}{4}\left(x^{1/2}\right)^2+b\left(x^{3/2}\right)^2+ c\frac{9}{16}\left(x^{-1/2}\right)^2\right]=\infty\,,
\end{equation}
we see that $\norm{\psi}=\infty$; and the mode is indeed discarded. Nonetheless a four-derivative norm will still fail to eliminate other non-analytic modes. To ensure that one properly removes all non-analytic modes one is forced to consider a Sobolev norm containing an infinite number of derivatives \cite{Warnick:2013hba}. 

In summary, the problem of computing the pseudospectrum of the QNMs seems to be ill-posed. To define the quasinormal modes we need to consider a Sobolev norm with infinite derivatives while the physically relevant norm is the energy which contains only two derivatives \cite{Boyanov:2023qqf}. More concretely, the energy norm fails to differentiate between modes entering and exiting the horizon for large enough $-\Im(\omega)$, i.e. for the overtones (in our convention exponential decay in time means $\Im(\omega)<0)$.

This interpretation then sheds light on the apparent reason as to why the pseudospectrum computed numerically fails to converge. In the numerical approach, the pseudospectrum is computed for a discretized version of the system using a Chebysev grid. This in turn implies that only analytic functions enter the computation, since non-analytic functions cannot be represented to infinite accuracy in the grid. However, as we go to the continuum limit where the grid is removed, the non-analytic functions can be better and better approximated in the grid. Hence, we tend to the pseudospectrum expected for the continuum limit where every point in the complex frequency plane satisfying $\Im{\omega}<a$, with $a$ a constant dependent on the number of derivatives in the norm \cite{Warnick:2013hba}, becomes a quasinormal frequency and thus the norm of the resolvent tends to infinity.

This seems to be indicating a very interesting physical picture. As we allow the perturbations entering in the pseudospectrum to be more and more localized in the radial direction (by increasing the number of grid points) the size of the perturbation needed to displace a given QNF to any arbitrary point in the complex plane becomes smaller and smaller (recall definition \ref{def:Pseudo Definition 2} of $\varepsilon$-pseudospectrum). This seems to agree with the results of \cite{Boyanov:2023qqf}, where the authors showed that for a fixed perturbation the effect on the spectrum was independent of the grid size. Consequently, the continuum limit corresponds to allowing more and more localized perturbations. However, one expects general relativity to break down at small scales. Thus, in this picture the grid size acts as a cutoff and to go to the continuum limit we expect to need to include higher derivative corrections to the theory and, consequently, to the energy norm; thus obtaining a Sobolev norm. A similar conclusion was reached in \cite{Carballo:2024kbk} in terms of the response of fluctuations to sources localized on the horizon. Further aspects of the instability of the pseudospectrum were also discussed in \cite{ourreview}.


\section{Complex Momentum Modes}\label{sect:Complex momentum modes}
Similar to QNMs, complex momentum modes are defined as solutions to the linearized equations of motion in a fixed black brane background satisfying infalling boundary conditions on the event horizon and normalizable boundary conditions on the AdS boundary. However, while QNMs are eigenvalues of the killing vector associated with time translations, complex \CMMs are eigenvalues of the generator of spatial translations along an arbitrary spatial direction parallel to the brane which we denote $x^3$. 

We consider a Schwarzschild anti-de Sitter black brane background with metric
\begin{equation}\label{eq:blackbrane}
    ds^2 =  \frac{r^2}{ l^2} (-f(r) d\hat{t}^2 + d\bold{x}^2)+\frac{l^2}{r^2 f(r)}dr^2  \,,\qquad f=1-\frac{r_h^4}{r^4}\,,
\end{equation}
where $f$ is the blackening factor which has a zero at $r=r_h$ and $l$ the AdS scale. We introduce regular coordinates \cite{Warnick:2013hba,Arean:2023ejh} by the following coordinate transformation 
\begin{equation}\label{eq:Regular Coordinates}
    t=\hat{t}-\frac{l^2}{r_h}\left(1-\frac{r_h}{r}\right)+\int \frac{dr}{f(r)}\left(\frac{l}{r}\right)^2\,, \qquad\rho=1-\frac{r_h}{r}\,,
\end{equation}
where we also have compactified the radius so that the AdS boundary is at $\rho=1$ and the horizon at $\rho=0$. Remarkably, on the boundary $t=\hat{t}$, which can be identified with the time in the dual field theory. The metric in regular coordinates takes the form
\begin{equation}\label{eq:GenericBgMetric_regular}
    ds^2=\frac{l^2}{z_h^2(1-\rho)^2}\left\{-f(\rho)dt^2+(dx^3)^2+d\bold{x}_\perp^2+2(1-f(\rho))z_hdtd\rho+(2-f(\rho))z_h^2d\rho^2\right\} \,,
\end{equation}
Here $z_h=l^2/r_h$, the Hawking temperature is given by $T=(\pi z_h)^{-1}$ and $\bold{x}_\perp$ denotes the spatial directions perpendicular to $x^3$ and $\rho$.
We note that regularity of the \CMMs is equivalent to demanding infalling boundary conditions.\footnote{Regular coordinates are the AdS analogue to the hyperboloidal ones used in asymptotically flat spacetimes \cite{Schmidt:1993rcx,Zenginoglu:2011jz,Ansorg:2016ztf,PanossoMacedo:2018hab,Bizon:2020qnd,PanossoMacedo:2023qzp}.} Alternatively one could have also used infalling Eddington-Finkelstein coordinates as in \cite{Cownden:2023dam}.

For a \CMM $\Phi(t,x^3,\bold{x}_\perp,z)$, after Fourier transforming along the time direction $t$ and the spatial coordinates $\bold{x}_\perp$
\begin{equation}
    \Phi(t,x^3,\bold{x}_\perp,z)=\int \frac{d\omega \,d^{\text{d}-2}k_\perp}{(2\pi)^{d-1}}\Phi(\omega,x^3,\bold{x}_\perp,z)\exp{-i\omega t+i \bold{k}_\perp \bold{x}_\perp}\,,
\end{equation}
the corresponding \CLM $k$ is defined through the following equation
\begin{equation}
    \partial_3 \Phi(\omega,x^3,\bold{x}_\perp,z)=i k \Phi(\omega,x^3,\bold{x}_\perp,z)\,,
\end{equation}
where note that $k$ is a function of $\omega$ and $\bold{k}_\perp$. Henceforth, we always take $\bold{k}_\perp=0$ as we can use the rotational invariance of the background \eqref{eq:GenericBgMetric_regular} to align the momentum with the $x^3$ direction. 

QNFs and \CLMs are just two different real sections of the complex lines that are the poles of the holographic retarded Green's function $G_R(\omega,k)$ in the two dimensional complex space spanned by $(\omega,k)$. 
QNFs are the sections $\Im(k)=0$ whereas \CLMs are the sections $\Im(\omega)=0$. Generically these complex lines intersect in complicated ways. That can make the properties of the modes very different when viewed as quasinormal modes or as complex momentum modes (see for example the discussion of the diffusive mode in \cite{Amado:2007pv} and its relevance for causality \cite{Landsteiner:2012gn}). It also gives rise to a rich and complex life story \cite{Grozdanov:2019uhi}.

\begin{figure}
\centering
    \begin{subfigure}{0.4\textwidth}
        \centering
        \begin{tikzpicture}

\definecolor{darkblue}{RGB}{30, 30, 180}
\definecolor{darkorange}{RGB}{255, 140, 0}

\draw[->] (-3, 0) -- (3, 0) node[right] {$\text{Re}(k)$};
\draw[->] (0, -3) -- (0, 3) node[above] {$\text{Im}(k)$};

\draw[ultra thick, darkblue, domain=0:180,dashed] plot ({2.75*cos(\x)}, {-2.75*sin(\x)});
\draw[ultra thick, darkblue,dashed] (2.75, 0) -- (-2.75, 0);

\draw[->, ultra thick, darkblue, >=stealth ] (1.94, -1.94) arc[start angle=-43, end angle=-45, radius=2.75];
\draw[->,  ultra thick, darkblue, >=stealth] (-1.94, -1.94) arc[start angle=-133, end angle=-135, radius=2.75];
\draw[->,   ultra thick, darkblue, >=stealth] (-1.27, 0) -- (-1.25, 0);
\draw[->, ultra thick, darkblue, >=stealth] (1.23, 0) -- (1.25, 0);

\foreach \i in {-1.75, -1.5, -1.25, -1,-0.75, -0.5, -0.25, 0.25, 0.5, 0.75,1, 1.25, 1.5 , 1.75} {
    \fill[red] (\i, \i) circle (2pt);
}

\node[darkblue] at (2.2, -2.4) {$\Gamma_{x^3<0}$};

\end{tikzpicture}
        \caption{}
        \label{fig:Lower}
    \end{subfigure}
    \hfill
    \begin{subfigure}{0.4\textwidth}
        \centering
        \begin{tikzpicture}

\definecolor{darkblue}{RGB}{30, 30, 180}
\definecolor{darkorange}{RGB}{255, 140, 0}

\draw[->] (-3, 0) -- (3, 0) node[right] {$\text{Re}(k)$};
\draw[->] (0, -3) -- (0, 3) node[above] {$\text{Im}(k)$};

\draw[ultra thick, darkblue, domain=0:180,dashed] plot ({2.75*cos(\x)}, {2.75*sin(\x)});
\draw[ultra thick, darkblue, dashed] (2.75, 0) -- (-2.75, 0);

\draw[->, ultra thick, darkblue, >=stealth] (1.94, 1.94) arc[start angle=43, end angle=45, radius=2.75];
\draw[->,  ultra thick, darkblue, >=stealth] (-1.94, 1.94) arc[start angle=132, end angle=135, radius=2.75];
\draw[->,  ultra thick, darkblue, >=stealth] (-1.27, 0) -- (-1.25, 0);
\draw[->, ultra thick, darkblue,,>=stealth] (1.23, 0) -- (1.25, 0);

\foreach \i in {-1.75, -1.5, -1.25, -1,-0.75, -0.5, -0.25, 0.25, 0.5, 0.75,1, 1.25, 1.5 , 1.75} {
    \fill[red] (\i, \i) circle (2pt);
}

\node[darkblue] at (2.2, 2.4) {$\Gamma_{x^3>0}$};

\end{tikzpicture}
        \caption{}
        \label{fig:Upper}
    \end{subfigure}
    \caption{The relevant integration contours for the poles in the complexified momentum-plane. Figure \ref{fig:Lower} shows the contour for $x < 0$, and figure \ref{fig:Upper}  shows the contour for $x > 0$. In order to obtain exponentially decaying waves travelling away from the origin of the perturbation, it is necessary that the poles lie in the 1st and 3rd quadrants for positive frequency (as assumed in the figure).}
    \label{fig:Contours}
    
\end{figure}
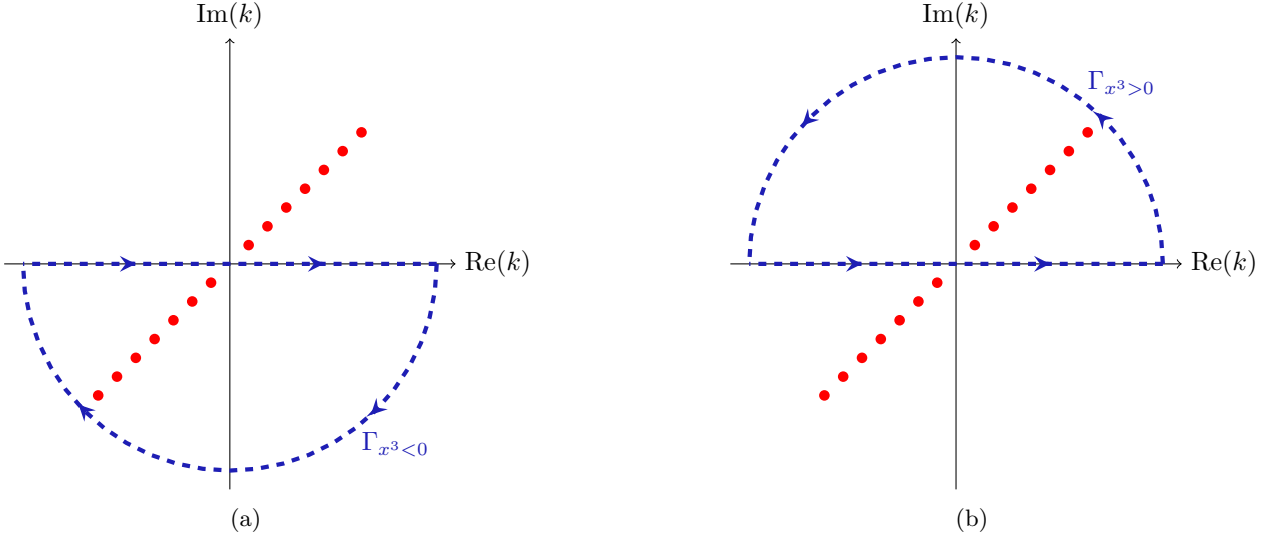

\CLMs and QNFs arise naturally in very different physical settings. Let us illustrate this from the point of view of the dual CFT. We consider a simple example within linear response theory and follow closely the discussion of \cite{Amado:2007pv}. In linear response theory the expectation value of an operator $\mathcal{O}$ is given by
\begin{equation}
    \expval{\mathcal{O}(t,\bold{x})}=-\int dt' \,d^{\text{d}-1}x'\,\, G_R(t-t',\bold{x}-\bold{x}')j_{\mathcal{O}}(t',x')\,,
\end{equation}
where $G_R$ is the retarded propagator and $j_{\mathcal{O}}$ the source of the said operator. If we now take the source to be an antenna of the form $j_{\mathcal{O}}=\chi(x^3)\exp(-i\omega t)$ with $\chi(x^3)$ an arbitrary function with analytic Fourier transform, then we get
\begin{align}
    \expval{\mathcal{O}(t,\bold{x})}&=-\int dt' \,d^{\text{d}-1}x'\,\, \int \frac{d\nu\, d^{d-1}q}{(2\pi)^{\text{d}}} \tilde{G}_R(\nu,\bold{q})\int \frac{dk}{2\pi}\tilde{\chi}(k)\exp{ik(x^3)'+iq(x-x')-i\omega t-i\nu (t-t'))}\nonumber\\
    &=\int_{-\infty}^{\infty} \frac{dk}{2\pi}\tilde{\chi}(k) \tilde{G}_R(\omega,k,\bold{k}_\perp=0)\exp(ikx^3-i\omega t)\nonumber\\
    &= -i\,\text{sign}(x^3)e^{-i\omega t}\sum_{k_n:\text{poles}} e^{ik_nx^3}\text{Res}\left[ \tilde{\chi}(k) \tilde{G}_R(\omega,k,\bold{k}_\perp=0)\right]\,,
\end{align}
where in the last equality we have used Cauchy's theorem (see figure \ref{fig:Contours}) and we have denoted the Fourier transforms with a tilde for clarity. Exponential decay away from the origin of the perturbation happens as long as
\begin{equation}\label{eq:cmmstatestability}
    \text{sign}\left(\frac{\omega}{\Re(k)} \right) = \text{sign}\left( \Im(k) \right)\,,
\end{equation}
which means that for positive frequency the poles are located in the first and third quadrant \cite{Amado:2007pv}.
    
Alternatively, we could take the source to be a hammer of the form $j_{\mathcal{O}}=\xi(t)\exp(i k x^3)$, with $\xi(t)$ a function with analytic Fourier transform. Then
\begin{align}
    \expval{\mathcal{O}(t,\bold{x})}&=-\int dt' \,d^{\text{d}-1}x'\,\, \int \frac{d\nu\, d^{d-1}q}{(2\pi)^{\text{d}}} \tilde{G}_R(\nu,\bold{q})\int \frac{d\omega}{2\pi}\tilde{\xi}(\omega)\exp{-i\omega t'+ik(x^3)'+iq(x-x')-i\nu (t-t'))}\nonumber\\
    &=\int_{-\infty}^{\infty} \frac{d\omega}{2\pi}\tilde{\xi}(\omega) \tilde{G}_R(\omega,k,\bold{k}_\perp=0)\exp{ikx^3-i\omega t}\nonumber\\
    &= i\Theta(t)\,e^{ik x^3}\sum_{\omega_n:\text{poles}} e^{-i\omega_nt}\text{Res}\left[ \tilde{\xi}(\omega) \tilde{G}_R(\omega,k,\bold{k}_\perp=0)\right]\,,
\end{align}

Thus, we conclude that \CLMs arise naturally when considering sources that behave as antennas, while QNFs appear when studying sources that behave as hammers. This has many deep fundamental implications as both setups are qualitatively very different. In particular, in the former case we always necessarily have a non-conservative system, even in the absence of thermalization, as the source introduces/extracts energy for any $t$; while in the latter setup we need not expect such behaviour as sources should vanish for large enough times $|t|>|t_*|$.

It is also worth stressing that when studying \CLMs $k(\omega)$ we are probing the stability of a sector with a given fixed frequency. This is particularly important in the $\omega=0$ limit. 
In this limit the \CMMs at $\omega=0$ are dual to the glueball masses of the effective theory of the lowest lying Kaluza-Klein modes of the theory compactifid on the thermal circle \cite{Witten:1998zw,Sakai:2004cn}. As this effective theory is conservative, we expect the spectrum of glueball masses to be spectrally stable. 
Hence, the \CLMs $k(\omega=0)$ should be stable provided we choose the norm adequately. On the other hand, a QNF $\omega=0$ need not be stable as we work in the fixed $k$ sector instead.

We briefly review the idea behind the interpretation of complex momenta as glueball masses in a dimensionally reduced theory. In order to do so we go momentarily back to the coordinate system \eqref{eq:blackbrane} but note that the time like Killing vectors of this metric and of \eqref{eq:GenericBgMetric_regular} are the same $\frac{\partial}{\partial \hat{t}} = \frac{\partial}{\partial t}$. In particular they have the same zero modes. In the black brane metric \eqref{eq:blackbrane} we can now do a double Wick rotation $\hat{t} \rightarrow i \sigma$ and $x^3\rightarrow i x^0$ The dual field theory lives then on the geometry $S^1\times M_3$ with $M_3$ being three dimensional Minkowski space. Zero modes on the $S^1$ are the zero modes of $\frac{\partial}{\partial_\sigma}$ which coincide with the zero modes of $\frac{\partial}{\partial \hat{t}}$. These zero modes live in three dimensional Minkowski space. The eigenvalues of the three dimensional momenta $k^2$ can be interpreted as masses of excitation in the dimensionally reduced theory (glueballs). Note that now $k^2<0$ simply means that $k$ is timelike in the reduced theory. Thus in the original theory, $\sqrt{-k^2}$ is interpreted as an absorption coefficient or inverse screening length, whereas in the double Wick rotated dimensionally reduced theory $\sqrt{-k^2}=M$ is interpreted as the mass of a propagating excitation of a (Hermitian) dual quantum field theory. 

Lastly we note that, the behaviour of \CMMs near the event horizon depends only on the frequency $\omega$. Hence, following the discussion on the previous section, we expect the pseudospectrum of \CMMs to converge as modes exiting the brane with $\Im(\omega)=0$ are discarded as we see in section \ref{subsect:ConstructionL}. We also explicitly see in section \ref{sect:Results} that our numerics do converge.

\section{Holographic model}\label{sect:Holographic model}
We want to study the spectral stability of the \CLMs of a real scalar field $\phi$ in a SAdS$_{4+1}$ background with metric \eqref{eq:GenericBgMetric_regular}. The corresponding action is given by
\begin{equation}\label{eq:Action scalar}
    S[\phi]=-\frac{1}{2}\int d^5x\left[ (\partial\phi)^2+m^2\phi^2\right]\,,
\end{equation}
where $m$ is the mass, which we take to be above the Breitenlohner-Freedman (BF) bound ($m^2l^2>-4$). We further assume standard quantization of the scalar, such that the leading mode in the asymptotic expansion around the AdS boundary $\rho=1$ is always the source term. 

Our stability analysis is based on the study of pseudospectra as discussed in section \ref{sect:ReviewPseudo}. In subsection \ref{subsect:EnergyNorm} we construct the energy norm and define the relevant function space. Following that, in subsection \ref{subsect:ConstructionL} we construct the eigenvalue problem and identify the relevant operator $\mathcal{L}$ and its adjoint $\mathcal{L}^\dagger$ with respect to the energy norm.

\subsection{Energy Norm and Function Space}\label{subsect:EnergyNorm}
Following \cite{Jaramillo:2020tuu} we try to define a function space which automatically imposes the adequate boundary conditions for the \CMMs and equip it with a physically-motivated norm based on the energy, which we label the energy norm.

We begin by first constructing the energy. To do so, recall that the energy momentum tensor for a real scalar field $\phi$ with action \eqref{eq:Action scalar} is given by
\begin{equation}
    T_{MN}[\phi]=\partial_M\phi\partial_N\phi-\frac{1}{2}g_{MN}\left[ (\partial\phi)^2+m^2\phi^2\right]\,,
\end{equation}
and that, consequently, the energy along a constant $t$-hypersurface ($\Sigma_t$) is
\begin{align}
    E[\phi]&= \int_{\Sigma_t} d^{d-1}x \sqrt{\gamma_\Sigma} \:n_\Sigma^M     T_{MN} \mathrm{t}^N   =-\int d\rho dx^1dx^2dx^3 \sqrt{-g}\:\mathrm{t}^M T^{t}\mathstrut_M\nonumber\\
    &=\int \frac{d\rho}{(1-\rho)^3}dx^1dx^2dx^3 \left\{f(\partial_\rho\phi)^2+z_h^2(\partial_a\phi)^2+z_h^2(\partial_3\phi)^2+z_h^2(2-f)(\partial_t\phi)^2+\frac{m^2l^2}{(1-\rho)^2}\phi^2\right\}\,,
\end{align}
where $\mathrm{t}=\partial_t$ is the killing vector associated with time translations, $n_\Sigma$ is the normal vector orthogonal to the integration surface and $\gamma_\Sigma$ the induced metric on it.  

Now, if we want to promote the expression for the energy above to a norm for the function space containing the \CMMs, we find ourselves facing two main problems. First, we would like to remove the $x^3$ integral so that the norm would not involve integration over all possible values of $x^3$, i.e., so that we do not need to know the explicit $x^3$ dependence of all $\phi$ in our function space. This is easily solved by considering the energy density along $x^3$
\begin{align}
    \varrho_E[\phi]&= \int_{\Sigma^\prime} d^{d-2}x \sqrt{\gamma_{\Sigma^\prime}} \:n_{(\Sigma^\prime)}{}^{MN} \mathrm{k}^{(3)}_M T_{NP} \mathrm{t}^P\nonumber\\
    &=\int \frac{d\rho}{(1-\rho)^3}dx^1dx^2 \left\{f(\partial_\rho\phi)^2+z_h^2(\partial_a\phi)^2+z_h^2(\partial_3\phi)^2+z_h^2(2-f)(\partial_t\phi)^2+\frac{m^2l^2}{(1-\rho)^2}\phi^2\right\}\,,
\end{align}
where $\Sigma^\prime$ is a hypersurface with $t$ and $x^3$ constant, $n_{(\Sigma^\prime)}$ the corresponding normal two-vector and $\mathrm{k}^{(3)}=k^{(3)\: M} \partial_M =\partial_3$. 

Second, we want to add an integral over time to be able to eliminate the dependence on time derivatives and focus on sectors with fixed frequency $\omega$. This can be achieved by considering a time-averaged energy density $\bar{\varrho}_E[\phi]$
\begin{equation}
    \bar{\varrho}_E[\phi]=\underset{s\rightarrow \infty}{\lim}\int_{-s}^s\frac{dt}{2s}\int \frac{d\rho dx^1dx^2}{(1-\rho)^3} \left\{f(\partial_\rho\phi)^2+z_h^2(\partial_a\phi)^2+z_h^2(\partial_3\phi)^2+z_h^2(2-f)(\partial_t\phi)^2+\frac{m^2l^2}{(1-\rho)^2}\phi^2\right\}\,.
\end{equation}

With all this, we can now Fourier transform in the directions $\bold{x}_\perp$ and $t$ and rewrite the expression above as 
\begin{equation}
     \bar{\varrho}_E[\phi]=\underset{s\rightarrow \infty}{\lim}\int\frac{d\omega d^2k_a}{2s(2\pi)^3}\frac{d\rho}{(1-\rho)^3} \left\{f\partial_\rho\phi\partial_\rho\bar{\phi}+z_h^2 (k_a)^2 \phi \bar{\phi}+z_h^2\partial_3\phi\partial_3\bar{\phi}+z_h^2(2-f)\omega^2 \phi\bar{\phi}+\frac{m^2l^2}{(1-\rho)^2}\phi\bar{\phi}\right\}\,,
\end{equation}
where we abuse notation and write $\phi=\phi(\omega,k_a,x^3,\rho)$ and an integral in $\omega$ as opposed the sum corresponding to finite $s$. We now consider a sector with fixed $\{\omega,\bold{k}_\perp\}$ and drop the integral over $\{\omega,\bold{k}_\perp\}$ as it only contributes with a prefactor which does not affect the operator norm. We further simplify by using the rotational invariance to set $\bold{k}_\perp=0$. Hence, the final expression for the time-averaged energy density without numerical prefactors is given by
\begin{equation}\label{eq:urnorm}
     \bar{\varrho}_E[\phi]=\int\frac{d\rho}{(1-\rho)^3} \left\{f\partial_\rho\phi\partial_\rho\bar{\phi}+z_h^2\partial_3\phi\partial_3\bar{\phi}+z_h^2(2-f)\omega^2 \phi\bar{\phi}+\frac{m^2l^2}{(1-\rho)^2}\phi\bar{\phi}\right\}\,.
\end{equation}

Now we promote the above expression to a physically-motivated inner product for studying \CMMs. As stated in section \ref{sect:Complex momentum modes}, these modes are regular (infalling) functions satisfying normalizable boundary conditions in the boundary of AdS which are solutions to the following eigenvalue problem
\begin{equation}
    z_h\partial_3 \phi=iz_hk\phi=i\mathfrak{q}\phi\,.
\end{equation}
where we have defined the dimensionless momentum $\mathfrak{q}$. Thus, we need to eliminate the explicit dependence on $\partial_3\phi$ from the inner-product to make it well defined. Mathematically, this necessity arises from the fact that the function space contains more elements than just the eigenvectors of $\partial_3$ (e.g. the sum of two eigenvectors with different eigenvalues is not an eigenvector). To solve this issue, we introduce an auxiliary field $\psi=z_h\partial_3\phi$, and arrive at the following expression for the energy 
norm on the space of doublets $\Psi=(\phi,\psi)$
\begin{equation}\label{eq:norm}
||\Psi||_E
=\int\frac{d\rho}{(1-\rho)^3} \left\{f\partial_\rho\bar{\phi}\partial_\rho\phi+\bar{\psi}\psi+\mathfrak{w}^2 (2-f)\bar{\phi}\phi+\frac{m^2l^2}{(1-\rho)^2}\bar{\phi}\phi\right\}\,,
\end{equation}
where $\mathfrak{w}=z_h\omega$.
To ensure convergence at $\rho=1$ we demand that $\phi=(1-\rho)^2 \xi(\rho)$ where $\xi(\rho)$ fulfills a Dirichlet boundary condition $\xi(1)=0$. The second and third terms are obviously positive definite. Expressing the kinetic and mass terms in the rescaled field we get
\begin{align}
&\int \frac{d\rho}{1-\rho} \left[ 
 (4f+m^2l^2) \xi^2 + 2 f \partial_\rho|\xi^2| + (1-\rho)^2 |\partial_\rho \xi|^2
 \right] = \nonumber\\
 &= \int \frac{d\rho}{1-\rho} \left[
 (4+m^2l^2 +4(1-\rho)^4)\,|\xi|^2+(1-\rho)^2 |\partial_\rho \xi|^2
 \right]   > 0 \,,
\end{align}
where we have done a partial integration and used that $f \xi^2$ vanishes on the horizon and on the boundary. The remainder is positive definite as long as $m^2l^2\geq-4$
and so is the norm.

We can further extend the norm to an inner product
\begin{equation}\label{eq:InnerProduct}
\expval{ \Psi_1 , \Psi_2 }_E
=\int\frac{d\rho}{(1-\rho)^3} \left\{f\partial_\rho\bar{\phi}_1\partial_\rho\phi_2+\bar{\psi}_1\psi_2+\mathfrak{w}^2 (2-f)\bar{\phi}_1\phi_2+\frac{m^2l^2}{(1-\rho)^2}\bar{\phi}_1\phi_2\right\}\,.
\end{equation}

In the next section we will show that only the \CMMs with infalling boundary conditions and normalizability at the AdS boundary belong to the Hilbert space defined by \eqref{eq:InnerProduct}.

\subsection{Construction of $\mathcal{L}$ and $\mathcal{L}^\dagger$}\label{subsect:ConstructionL}
We now want to use the equation of motion of the scalar field \eqref{eq:Action scalar}
\begin{equation}\label{eq:ScalarEoM}
    \nabla^2\phi-m^2\phi=0\,,
\end{equation}
to rewrite the eigenvalue equation 
\begin{equation}
    z_h\partial_3 \Psi =i\mathfrak{q}\Psi\,.
\end{equation}
in terms of a differential operator $\mathcal{L}=\mathcal{L}[\mathfrak{w},\rho;\partial_\rho,\partial_\rho^2]$. We achieve this by first explicitly writing the equation of motion \eqref{eq:ScalarEoM} as
\begin{equation}
       z_h^2\partial_3^2\phi=\frac{m^2l^2}{(1-\rho)^2}\phi-(1-\rho)^3 \partial_\rho\left(\frac{f\partial_\rho\phi}{(1-\rho)^3}\right)+\mathfrak{w}^2(-2+f)\phi+i\mathfrak{w}\left[(1-\rho)^3\partial_\rho\left(\frac{(1-f)\phi}{(1-\rho)^3}\right)+(1-f)\partial_\rho\phi  \right]\,.
\end{equation}
and then introducing the auxiliary field $\psi=z_h\partial_3\phi$ so that we get
\begin{equation}\label{eq:Def L for doublet}
    i\mathfrak{q}\Psi=\begin{pmatrix}
        0&1\\L&0
    \end{pmatrix}\begin{pmatrix}
        \phi\\\psi
    \end{pmatrix}=\mathcal{L}\Psi \,,
\end{equation}
where $L$ is a differential operator whose action on $\phi$ is
\begin{equation}
    L\phi=\frac{m^2l^2}{(1-\rho)^2}\phi-(1-\rho)^3 \partial_\rho\left(\frac{f\partial_\rho\phi}{(1-\rho)^3}\right)+\mathfrak{w}^2(-2+f)\phi+i\mathfrak{w}\left[(1-\rho)^3\partial_\rho\left(\frac{(1-f)\phi}{(1-\rho)^3}\right)+(1-f)\partial_\rho\phi  \right]\,,
\end{equation}
and $\mathcal{L}=\mathcal{L}[\mathfrak{w},\rho;\partial_\rho,\partial_\rho^2]$ is the differential operator which we initially wanted to obtain.

We now argue that only the infalling  and normalizable solutions belong to the function space introduced in the previous section. First we note that the differential operator has regular singular points at $\rho=0$ and $\rho=1$. The coefficients of the derivatives are analytic in the interior $\rho\in(0,1)$. We can therefore construct two analytic solutions valid in the interior by a simple power series ansatz. Therefore solutions are certainly locally integrable except possibly for the boundary points. At the (conformal) boundary of AdS $(\rho=1)$ it is well known that the two local solution behave like
\begin{equation}
    \phi^{(\mathrm{nn})} \propto (1-\rho)^{\Delta_-}\,,\qquad \phi^{(\mathrm{n})} \propto  (1-\rho)^{\Delta+}\,,
\end{equation}
with $\Delta_{\pm} = \left( 2\pm\sqrt{4+m^2 l^2} \right)$. Integrability of the inner product demands $\Delta_a + \Delta_b > 4$ with $a,b\in\{+,-\}$. We see that in the range of masses $-4 < m^2 l^2\leq 0$ the inner product \eqref{eq:InnerProduct} indeed rejects the solution $\phi^{(\mathrm{nn})}$ as non-normalizable.\footnote{We restrict ourselves here to standard quantization.} 

Now let us proceed to the Horizon at $\rho=0$. There we have to distinguish between the cases $\omega\neq 0$ and $\omega=0$. For $\omega\in \mathbb{R}\backslash\{0\}$ the two local solutions near the horizon behave as
\begin{equation}
\phi^{(\mathrm{in})} \propto \rho^0\,,\qquad \phi^{(\mathrm{out})}\propto \rho^{i\mathfrak{w}/2}\,.
\end{equation}
The most singular term is the derivative term in the norm. Since the blackening factor $f\propto \rho$ near the Horizon we find
\begin{equation}
    \bar\varrho_E \approx \int d\rho \,\rho\, |\partial_\rho\phi^{(\mathrm{out})} |^2 \approx  \int d\rho\, \rho\, \frac{1}{\rho^2} \rightarrow \infty\,.
\end{equation}
For $\omega=0$ the local solutions behave as $\phi^{(\mathrm{in})}\propto \rho^0$ and $\phi^{(\mathrm{out})}\propto \log(\rho)$. Clearly the logarithmic solution again is rejected by the norm as non-integrable at the horizon. We see therefore that our choice of norm is such that only solutions to the eigenvalue problem with the correct boundary conditions (infalling at the horizon and normalizable at the boundary) are elements of the Hilbert space defined through the inner product \eqref{eq:InnerProduct}. There is a further subtlety concerning the asymptotic behavior of $\psi$. If we only demand that the norm exists then it seems enough to demand that $\psi$ vanishes faster than $(1-\rho)$ towards the boundary. However we also want $\norm{\mathcal{L} \Psi}$ to be well defined thus $\psi$ has to go faster that $(1-\rho)^2$ towards the boundary. 

The eigenvalues of $\mathcal{L}$ are the \CLMs times the imaginary unit $i$. In practice, when presenting our results in section \ref{sect:Results}, we find it more convenient to work with differential operator $\tilde{\mathcal{L}}=-i\mathcal{L}$ rather than with $\mathcal{L}$. This choice ensures that the eigenvalues are the \CLMs without any imaginary prefactor and does not alter any of the relevant conclusions derived throughout this section in terms of $\mathcal{L}$.

Regarding the adjoint $\mathcal{L}^\dagger$, we have the following expression 
\begin{align}
&\expval{\mathcal{L}^\dagger\Psi_1,\Psi_2}_E - \expval{\mathcal{L}\Psi_1,\Psi_2}_E=\nonumber\\
&=\int\frac{d\rho}{(1-\rho)^3}\Biggl\{\left[2\mathfrak{w}^2(2-f)\bar{\phi}_1\psi_2-i\mathfrak{w}(1-f)\phi_2\partial_\rho\bar{\psi}_1-i\mathfrak{w}(1-\rho)^3\phi_2\partial_\rho\left(\frac{(1-f)\bar{\psi}_1}{(1-\rho)^3}\right) - (\phi\leftrightarrow\psi) \right] \nonumber\\
&+\phi_2\left[-2i\mathfrak{w}\bar{\psi}_1\delta(\rho)\right]\Biggr\}\,.
\end{align}


Thus for generic $\mathfrak{w}^2\neq0$ we find that $\mathcal{L}^\dagger\neq \mathcal{L}$. Moreover, we cannot rewrite $\mathcal{L}^\dagger$ as $\mathcal{L}$ plus some boundary terms; we necessarily have contributions to the non-normality arising from the bulk. This nicely matches the discussion of section \ref{sect:Complex momentum modes}. We expect that the system is non-conservative even in the absence of temperature, hence there should be contributions to the non-normality that arise independently of the existence of an horizon, i.e. contributions arising from the bulk. In fact, as we see in appendix \ref{app:NonNormalityEmptyAdS}, the existence of bulk contributions for $\mathfrak{w}\neq0$ is guaranteed even in AdS, further confirming the physical intuition. 

Remarkably $\mathcal{L}$ is self-adjoint  at $\mathfrak{w}=0$ and has a stable spectrum. This is in particularly good agreement with the arguments presented in section \ref{sect:Complex momentum modes}. At $\mathfrak{w}=0$ we are studying the glueball spectrum of the dimensionally reduced theory and, as this theory lacks any dissipation mechanism, a well-defined norm should predict that $\mathcal{L}$ is Hermitian; which indeed matches our results. 

 \section{Numerical method}\label{sect:Numerical method}
We follow \cite{Jaramillo:2020tuu,Arean:2023ejh} and perform a numerical stability analysis using pseudospectra {\'a} la \cite{Trefethen:2005}. For completeness, here we briefly review the some of the details and subtleties. 

Our numerical approach is based on the discretization of the radial coordinate $\rho$ on a Chebyshev grid with points:
\begin{equation}\label{eq:Chebyshev grid}
    \rho_j=\frac{1}{2}\left[1-\cos\left(\frac{j\pi}{N}\right)\right]\,,\qquad j=0,1,...,N\,,
\end{equation}
and the discretization of the differential operators using the corresponding Chebyshev differentiation matrices \cite{Trefethen:2000}.

In order to numerically select the correct function space, we find it convenient work with the rescaled scalar doublet $u$ defined as
\begin{equation}
    u=(1-\rho)^2 \Psi\,,
\end{equation}
where we impose Dirichlet boundary conditions for $u$ at $\rho=1$. This boundary condition can be easily implemented by removing the rows and columns corresponding to the AdS boundary from all discretized operators, including the discretized energy norm. We use \textit{Wolfram Engine} to compute condition numbers and pseudospectra of the resulting matrices according to the discussion presented in section \ref{sect:ReviewPseudo}. 

\begin{figure}[htb!]
    \centering
        \includegraphics[width=0.6\linewidth]{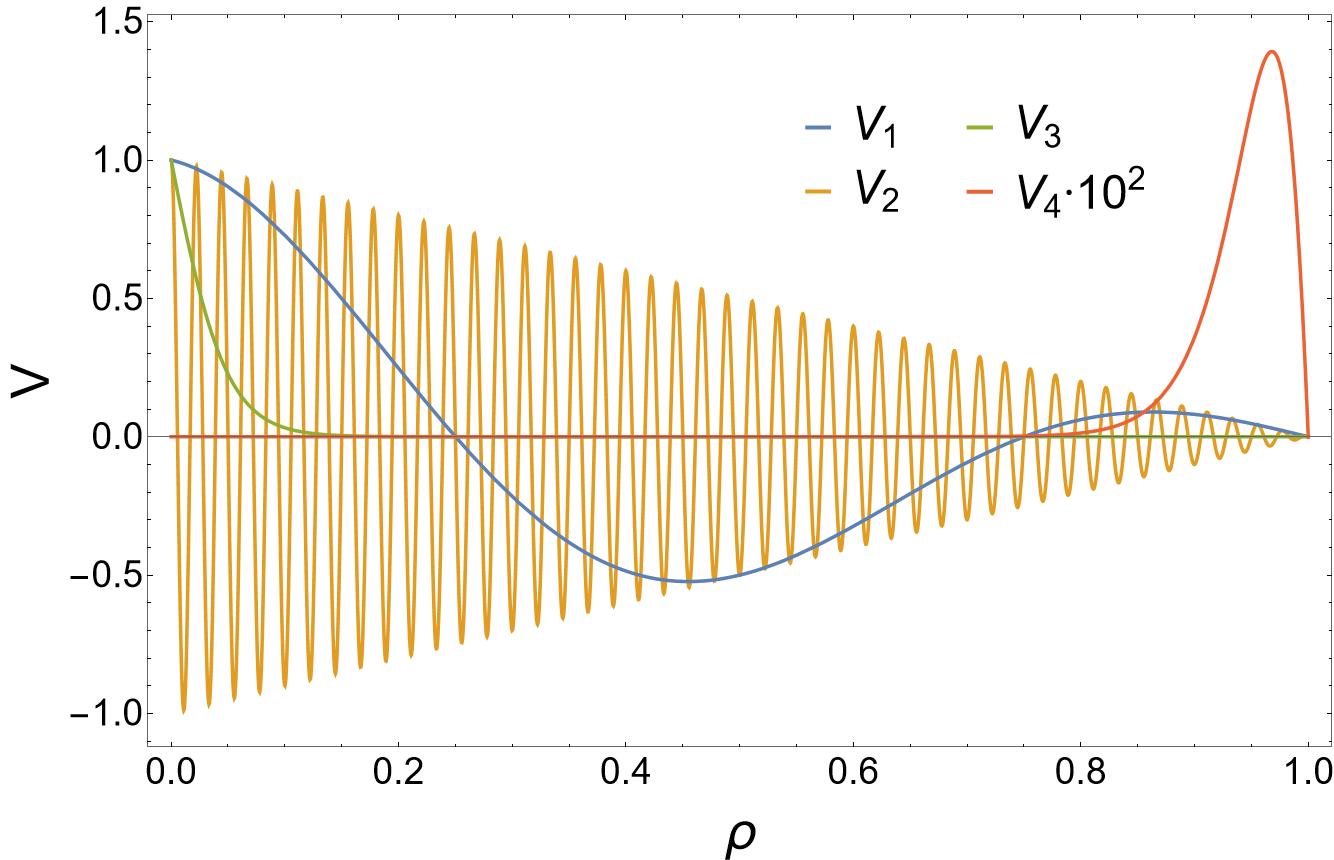}
      \caption{Deterministic potentials
      \eqref{eq:deterministicpots}
      with $A_i=1$.
      Recall that the horizon is at $\rho=0$ and the boundary at $\rho=1$.}
        \label{fig:Deterministic Potentials plot}
\end{figure}

To further explore the origin of the (in)stability, we study the selective pseudospectra associated with local potential perturbations to the original equation of motion \eqref{eq:ScalarEoM}. Concretely, we consider perturbed equations of motion of the form:
\begin{equation}\label{eq:eoms perturbed}
   \left[-\nabla_M \nabla^M + m^2 + \frac{V(\rho)}{l^2}\right]\phi=0\,,
\end{equation}
where, in order to preserve the asymptotic behavior on the AdS boundary, we choose potentials with $V(1)=0$. The selective pseudospectrum is then computed using randomly generated potential perturbations constructed as diagonal matrices with random entries and normalized to a given size. Moreover, we complement this analysis with the computation of the \CLMs for the perturbed system with the following deterministic potentials:
\begin{subequations}
\begin{align}
    V_1(\rho)&=A_1(1-\rho)\cos(2\pi\rho)\label{eq:determinisiticV1}\,,\\
    V_2(\rho)&=A_2(1-\rho)\cos(90\pi\rho)\label{eq:determinisiticV2}\,, \\
    V_3(\rho)&=A_3(1-\rho)\left\{1-\tanh\left[20\rho\right]\right\}\label{eq:determinisiticV3}\,,\\
    V_4(\rho)&=A_4(1-\rho)\left\{1-\tanh\left[20(1-\rho)\right]\right\}\label{eq:determinisiticV4}\,,
\end{align}\label{eq:deterministicpots}%
\end{subequations}  
which shed light on a few interesting regimes. With $V_1$ and $V_2$, we probe the effect of long and short $\rho$-wavelength (wavelength in the $\rho$ direction) perturbations; while with $V_3$ and $V_4$, we analyze the stability under localized perturbations near the horizon (IR of the dual CFT) and the boundary (UV of the dual CFT). The coefficients $A_i$ are normalization constants that fix the magnitude of the perturbation.
We plot these potentials in figure~\ref{fig:Deterministic Potentials plot}.

\section{Results}\label{sect:Results}

Here we present the results of the analysis described in the previous sections. Our numerical simulations are performed in a grid of 100 points with a precision of 5$\times$MachinePrecision. 

In appendix \ref{app:cmms} we some the values for the \CLMs at different masses $m^2l^2$ and frequencies $\mathfrak{w}$. All pseudospectrum plots are collected in appendix \ref{app:PseudoPlots}.

We are mainly interested in the stability of the \CLMs lying closest to the real axis as these dominate the long-distance behaviour of the system.\footnote{The system is symmetric under $\mathfrak{q}\rightarrow-\mathfrak{q}$. Consequently, henceforth when talking about any given \CLM we will be generically referring to the pair $\{\mathfrak{q},-\mathfrak{q}\}$.} For this reason our computations are restricted to the region of the complex $\mathfrak{q}$-plane defined by $|\Im(\mathfrak{q})|<10$ and $|\Re(\mathfrak{q})|<10$. To ensure the convergence of the pseudospectrum in that region, in figure \ref{fig:ConvergencePlot} we plot the value of the inverse of the norm of the resolvent $\norm{(\mathcal{L}-\mathfrak{q})^{-1}}$ at the edges of the said region as a function of $N$.
Clearly, we can see that in agreement with the discussion presented in section \ref{sect:ConvergencePseudoQNFs}, the pseudospectrum indeed converges for real $\mathfrak{w}$. 

In figures \ref{fig:CloseupPseudospectraScalar} and \ref{fig:LargePseudospectraScalar} we plot the full and selective pseudospectra and the corresponding condition numbers for different values of $m^2l^2$ and $\mathfrak{w}$. As expected from the analytic results discussed in section \ref{subsect:ConstructionL}, at zero frequency the spectrum is indeed stable as the operator becomes normal. We can easily appreciate this in figures \ref{fig:ScalarAdS5_CloseupPseudo_m0w0} and \ref{fig:ScalarAdS5_CloseupPseudo_mn3w0} where the $10^{-1}$ pseudospectrum around the first \CLM is shown to match the circle of radius $10^{-1}$ around the said momenta. At non-zero $\mathfrak{w}$ the full pseudospectrum opens up denoting spectral instability. Remarkably, at $\mathfrak{w}=10$ we find that in order to drive the background unstable by making one of complex momenta cross into the second or fourth quadrants of the complex plane \cite{Amado:2007pv}, we need to consider perturbations of size $0.7$ and $0.5$, for masses $m^2l^2=0$ and $m^2l^2=-3$, respectively. As such perturbations are of the order of magnitude of the distance between \CLMs, we consider that they are relatively big. Hence we claim that while the spectrum of \CLMs is unstable, this spectral instability cannot drive the background unstable. We refer the reader to section 3.3 of \cite{Arean:2023ejh} for a detailed discussion on the difference between spectral and background instability. It is also worth noting that, similarly to what happened for QNFs, higher \CMMs are increasingly unstable as indicated by their condition numbers (see figure \ref{fig:LargePseudospectraScalar}).

Regarding the selective pseudospectrum of local potential perturbations, we find a striking decrease in instability compared to the instability found for generic perturbations. In particular, we have not managed to find any random local potential perturbation capable of destabilizing the lowest lying \CLM (see figure \ref{fig:CloseupPseudospectraScalar}). This is in contrast with the observations for the lowest QNF in \cite{Arean:2023ejh}.
Furthermore, for higher \CLMs, we always find that the effect of these perturbations is much smaller than that of generic perturbations (see figure \ref{fig:LargePseudospectraScalar1stQ}). This suggests that in order to cause a significant impact on the spectrum, one necessarily has to consider non-local potentials. 

To further explore this surprising spectral stability, in figures \ref{fig:CloseupDeterministicScalar} and \ref{fig:LargeDeterministicScalar} we plot the effect of the deterministic potentials \eqref{eq:deterministicpots} on the lowest lying \CLM and on the full region of the complex plane under consideration, respectively. We see that as indicated by the selective pseudospectrum, the first \CLM is stable under the deterministic potentials \eqref{eq:deterministicpots}. Moreover, in figure \ref{fig:LargeDeterministicScalar} we find that this stability is shared by all the plotted \CLMs; as we observe that the perturbations do not displace them in a significant manner. This contrasts greatly with the picture found for the QNFs in \cite{Arean:2023ejh}, thus exemplifying the different stability properties of complex momenta and quasinormal frequencies in agreement with the discussion of section \ref{sect:Complex momentum modes}.

To conclude, in figures \ref{fig:mDependence} and \ref{fig:wDependence} we plot the condition numbers for the first three \CLMs as a function of $m^2l^2$ and $\mathfrak{w}$, respectively. As expected at $\mathfrak{w}=0$ the condition numbers are 1, up to small numerical errors. Remarkably we find that instability increases with mass and with frequency. From the dual perspective this implies that absorption lengths of operators with larger conformal dimension at large frequencies are more unstable.

\section{Summary and Discussion}\label{sect:Conclusions}
This work presents complex linear momenta (\CLMs) as a new probe to explore the spectral instability of holographic theories. From the gravitational point of view, these are eigenvalues of the translation operator along a space-like direction parallel to the AdS boundary at fixed, real frequency, subject to appropriate boundary conditions. From the holographic perspective, \CLMs are dual to the poles of the retarded Green's function at fixed real frequency. They contain information about the causal structure of the dual field theory and, at zero frequency, precisely describe the masses of the states in the theory compactified on the thermal circle.

The study of Complex Momentum Modes (\CMMs ), with non-zero frequency ( $\omega\neq 0$), implies pumping energy into the system; making it non-conservative and thus vulnerable to spectral instabilities. In particular, we have computed the pseudospectra of \CLMs for a massive, real scalar on a Schwarzschild-AdS$_{4+1}$ background. In order to measure the displacement of the momenta, we use the (two derivative) energy norm. Remarkably, we have found that \CLMs are unstable at non-zero frequency. Furthermore, we have argued that the spectral stability observed at zero frequency is a physical feature related to the fact that in that case \CLMs are dual to the glueball spectrum of a conservative theory arising from dimensional reduction on the thermal circle. At non-zero frequency, the spectral instability increases with mass and frequency. This suggests that in the dual field theory, the poles of retarded Green's functions of scalar operators with large mass dimension at large frequency are more unstable. Additionally, the pseudospectrum appears to be particularly resistant to local potential perturbations, indicating that non-local potentials may need to be considered to produce a significant displacement of the \CLM. 

As explained in Section \ref{sect:ConvergencePseudoQNFs} the QNF pseudospectrum fails to converge for negative enough $\Im(\omega)$. One of the paper's main results is the striking numerical convergence of the  \CLMs pseudospectra in our particular setup. We expect this to be a generic property since, for branes without momenta along the $x^3$ direction, the near-horizon behaviour of \CMMs is independent of $k$. Thus, we claim that the pseudospectrum of \CLMs offers a new alternative to assess the spectral stability of the theory, which allows us to derive quantitative results, which in turn would facilitate making precise predictions for observables.

Even though QNMs and \CMMs seem very similar in flavour, we have shown that they probe different regimes of the theory. As such, neither the stability of their pseudospectra nor their numerical convergence are directly related. It seems odd since both quantities are holographically dual to the same object, poles in the retarded Green's functions. We are looking at two different real sections of the solutions of $G_R(k, \omega )^{-1}=0$. In principle analytic continuation should allow us to relate both sections. If such an ``analytic continuation'' of sorts was possible, we could then attempt to reconstruct a numerically convergent pseudospectra for QNFs; without needing to employ higher derivative Sobolev norms. This, however, is well beyond the scope of this work.

\section*{Acknowledgements}
We thank D. Areán, V. Boyanov and J.L. Jaramillo for valuable discussions.
This work is supported through the grants CEX2020-001007-S and PID2021-123017NB-100, PID2021-127726NB-I00 funded by MCIN/AEI/10.13039/501100011033 and by ERDF ``A way of making Europe''. 
The work of D.G.F. is supported by FPI grant PRE2022-101810. 
The work of PSB is supported by Fundaci\'on S\'eneca, Agencia de Ciencia y Tecnolog\'ia de la Regi\'on de Murcia, grant 21609/FPI/21 and Spanish Ministerio de Ciencia e Innovación PID2021-125700NA-C22.

\appendix

\section{Non-normality in AdS}\label{app:NonNormalityEmptyAdS}
To better understand the bulk contributions to $\mathcal{L}^\dagger$ that we found in \ref{subsect:ConstructionL}, here we repeat the computations of the preceding sections taking the AdS limit $f=1$. As we are interested only in the bulk contributions, we do not keep track of possible boundary terms in the computation of $\mathcal{L}^\dagger$.  

In AdS$_{4+1}$, the operator $\mathcal{L}$ reads 
\begin{equation}
   \mathcal{L}_0=\begin{pmatrix}
        0&1\\L_0&0
    \end{pmatrix}\,,
\end{equation}
where $L_0$ is a differential operator whose action on $\phi$ is defined as
\begin{equation}
    L_0\phi=\frac{m^2l^2}{(1-\rho)^2}\phi-(1-\rho)^3 \partial_\rho\left(\frac{\partial_\rho\phi}{(1-\rho)^3}\right)-\mathfrak{w}^2\phi\,,
\end{equation}
and the energy norm simplifies to
\begin{equation}
\expval{\Psi_1,\Psi_2}_E=\int\frac{d\rho}{(1-\rho)^3} \left\{\partial_\rho\bar{\phi}_1\partial_\rho\phi_2+\bar{\psi}_1\psi_2+\mathfrak{w}^2\bar{\phi}_1\phi_2+\frac{m^2l^2}{(1-\rho)^2}\bar{\phi}_1\phi_2\right\}\,.
\end{equation}
Hence, a straightforward computation allows us to conclude the following 
\begin{align}
&\expval{\mathcal{L}_0^\dagger\Psi_1,\Psi_2}_E - \expval{\mathcal{L}_0\Psi_1,\Psi_2}_E=\int\frac{d\rho}{(1-\rho)^3}\left\{\left[2\mathfrak{w}^2\bar{\phi}_1\psi_2 - (\phi\leftrightarrow\psi) \right] \right\}\,.
\end{align}
which, as anticipated, shows that $\mathcal{L}_0^\dagger$ contains bulk contributions to the non-normality for $\mathfrak{w}\neq0$.

Remarkably, the existence of non-normality even in the absence of a black hole matches the naive physical intuition discussed in section \ref{sect:Complex momentum modes}. From the CFT perspective, exciting \CMMs corresponds to turning on a source of the form $j(t,x^3)=\chi(x^3)e^{-i\omega t}$ for the dual operator; thus, even without thermal dissipation, the system has an influx/outflow of matter associated with the time dependent source, which in turn makes the system non-conservative. Again, in agreement with the naive physical intuition, these bulk contributions also vanish in the $\mathfrak{w}=0$ limit, signaling that the system becomes conservative as the source is no longer time-dependent.

\section{Numerical Values of the Complex Momenta}\label{app:cmms}
In this appendix we provide the numerical values of the first 10 \CLMs for the real scalar with action \eqref{eq:Action scalar}. For purposes of presentation we limit the precision to 15 significant figures.

\hfill\break

\begin{table}[!htb]
    \centering
    \begin{tabular}{|c|c|c|}
        \hline
         $n$&$\Re(\mathfrak{q}_n)$&$\Im(\mathfrak{q}_n)$  \\
         \hline
         \hline
         $1$& $0$ & $\pm 3.40406555853814$ \\
         \hline
         $2$& $0$ & $\pm 5.87596663030813$ \\
         \hline
         $3$& $0$ & $\pm 8.30511662249606$ \\
         \hline
         $4$& $0$ & $\pm 10.7196286254632$ \\
         \hline
         $5$& $0$ & $\pm 13.1274967745451$ \\
         \hline
         $6$& $0$ & $\pm 15.5317934290101$ \\
         \hline
         $7$& $0$ & $\pm 17.9339496075144$ \\
         \hline
         $8$& $0$ & $\pm 20.3347218217190$ \\
         \hline
         $9$& $0$ & $\pm 22.7345476095668$ \\
         \hline
         $10$& $0$ & $\pm25.1336976323937$ \\
         \hline
    \end{tabular}
    \caption{\CLMs for $m^2l^2=0$ and $\mathfrak{w}=0$.}
\end{table}



\begin{table}[!htb]
    \centering
    \begin{tabular}{|c|c|c|}
        \hline
         $n$&$\Re(\mathfrak{q}_n)$&$\Im(\mathfrak{q}_n)$  \\
         \hline
         \hline
         $1$& $\pm 9.02226257097657$ & $\pm 1.87239874470858$ \\
         \hline
         $2$& $\pm 8.17892555723022$ & $\pm 3.97449830746049$ \\
         \hline
         $3$& $\pm 7.50532453968757$ & $\pm 6.40222333417884$ \\
         \hline
         $4$& $\pm 7.04748494704286$ & $\pm 8.96919759174839$ \\
         \hline
         $5$& $\pm 6.75242590366829$ & $\pm 11.5596060709257$ \\
         \hline
         $6$& $\pm 6.55981285259286$ & $\pm 14.1324913026753$ \\
         \hline
         $7$& $\pm 6.42974689165010$ & $\pm 16.6795521362831$ \\
         \hline
         $8$& $\pm 6.33868833659767$ & $\pm 19.2024829197097$ \\
         \hline
         $9$& $\pm 6.27280863252675$ & $\pm 21.7051740861311$ \\
         \hline
         $10$& $\pm 6.22376285335282$ & $\pm 24.1914304968759$ \\
         \hline
    \end{tabular}
    \caption{\CLMs for $m^2l^2=0$ and $\mathfrak{w}=10$.}
\end{table}

\begin{table}[!htb]
    \centering
    \begin{tabular}{|c|c|c|}
        \hline
         $n$&$\Re(\mathfrak{q}_n)$&$\Im(\mathfrak{q}_n)$  \\
         \hline
         \hline
         $1$& $0 $ & $\pm 2.39628046947118$ \\
         \hline
         $2$& $0 $ & $\pm 4.79256093894237$ \\
         \hline
         $3$& $0 $ & $\pm 7.18884140841355$ \\
         \hline
         $4$& $0 $ & $\pm 9.58512187788474$ \\
         \hline
         $5$& $0 $ & $\pm 11.9814023473559$ \\
         \hline
         $6$& $0 $ & $\pm 14.3776828168271$ \\
         \hline
         $7$& $0 $ & $\pm 16.7739632862983$ \\
         \hline
         $8$& $0 $ & $\pm 19.1702437557695$ \\
         \hline
         $9$& $0 $ & $\pm 21.5665242252407$ \\
         \hline
         $10$& $0$ & $\pm 23.9628046947118$ \\
         \hline
    \end{tabular}
    \caption{\CLMs for $m^2l^2=-3$ and $\mathfrak{w}=0$.}
\end{table}

\begin{table}[!htb]
    \centering
    \begin{tabular}{|c|c|c|}
        \hline
         $n$&$\Re(\mathfrak{q}_n)$&$\Im(\mathfrak{q}_n)$  \\
         \hline
         \hline
         $1$& $\pm 9.38974080989012$ & $\pm 1.10875947814635$ \\
         \hline
         $2$& $\pm 8.54763762913947$ & $\pm 2.94898874493230$ \\
         \hline
         $3$& $\pm 7.78565832114999$ & $\pm 5.22055753548535$ \\
         \hline
         $4$& $\pm 7.23018023647193$ & $\pm 7.72858499087675$ \\
         \hline
         $5$& $\pm 6.86711966426586$ & $\pm 10.3091717841825$ \\
         \hline
         $6$& $\pm 6.63309570199478$ & $\pm 12.8879323622292$ \\
         \hline
         $7$& $\pm 6.47813029086701$ & $\pm 15.4439100354085$ \\
         \hline
         $8$& $\pm 6.37172016550565$ & $\pm 17.9750953890531$ \\
         \hline
         $9$& $\pm 6.29604904918147$ & $\pm 20.4845878580473$ \\
         \hline
         $10$& $\pm 6.24054508579031$ & $\pm 22.9762411905281$ \\
         \hline
    \end{tabular}
    \caption{\CLMs for $m^2l^2=-3$ and $\mathfrak{w}=10$.}
\end{table}

\clearpage

\section{Pseudospectrum plots}\label{app:PseudoPlots}
In this appendix we collect the pseudospectrum plots discussed in section \ref{sect:Results}. 

\hfill\break

\begin{figure}[h]
    \centering
    \begin{subfigure}[b]{0.49\linewidth}
        \centering
        \includegraphics[width=.935\linewidth]{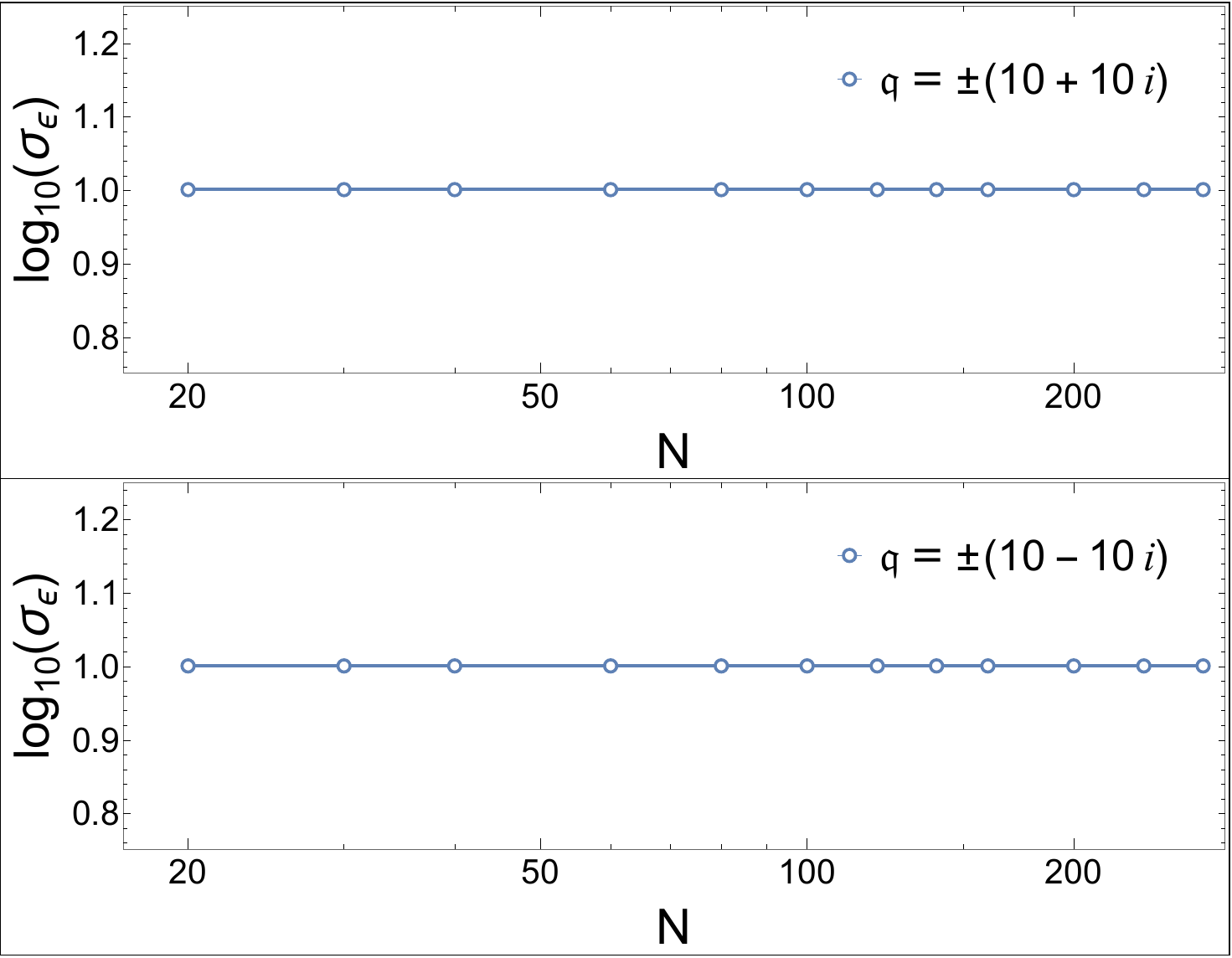}
        \captionsetup{justification=centering}
        \caption{$m^2l^2=0$, $\mathfrak{w}=0$.}
        \label{fig:Convergencem0w0}
    \end{subfigure}\hfill
    \begin{subfigure}[b]{0.49\linewidth}
        \centering
        \includegraphics[width=.935\linewidth]{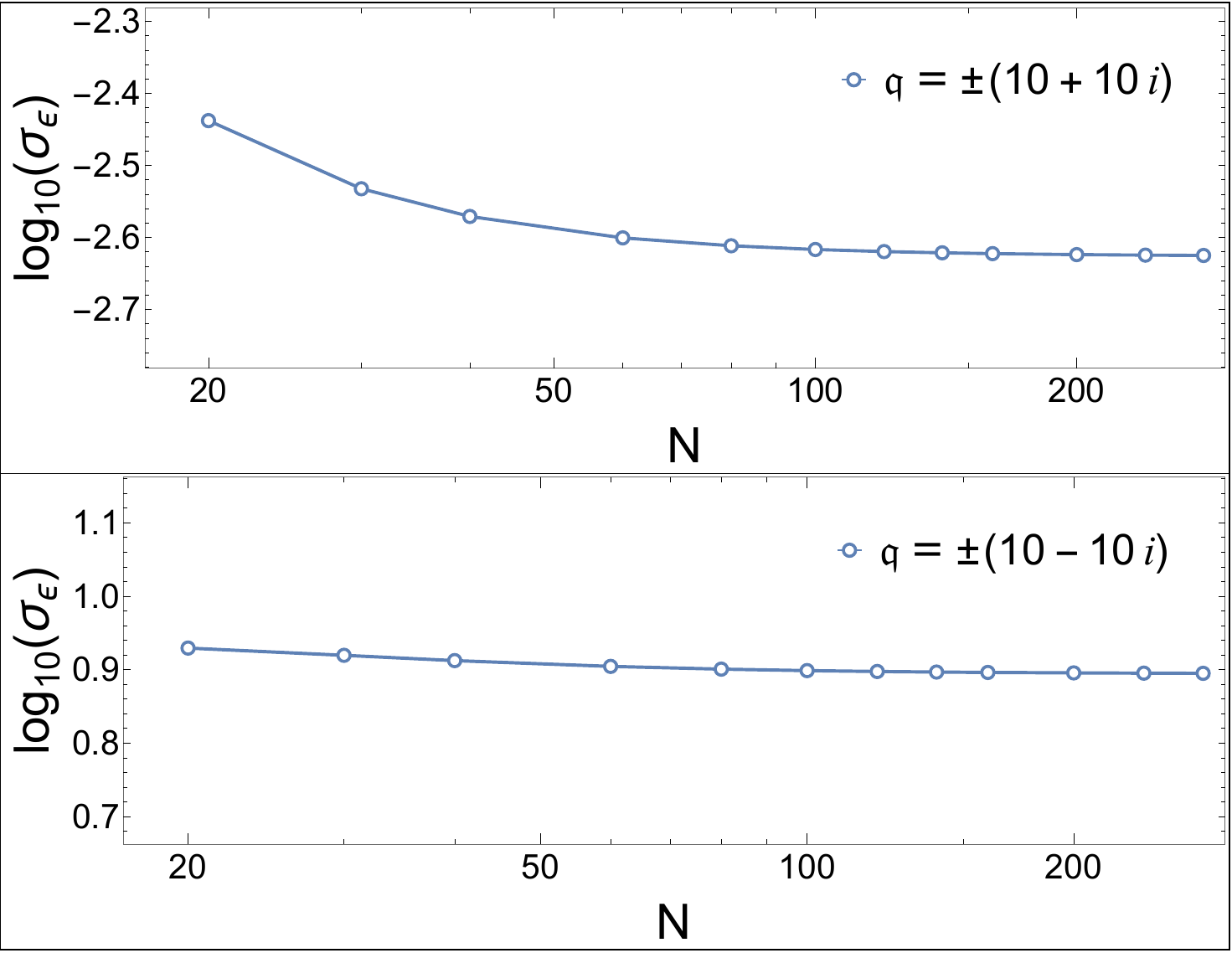}
        \captionsetup{justification=centering}
        \caption{$m^2l^2=0$, $\mathfrak{w}=10$.}
        \label{fig:Convergencem0w10}
    \end{subfigure}
        \begin{subfigure}[b]{0.49\linewidth}
        \centering
        \includegraphics[width=.935\linewidth]{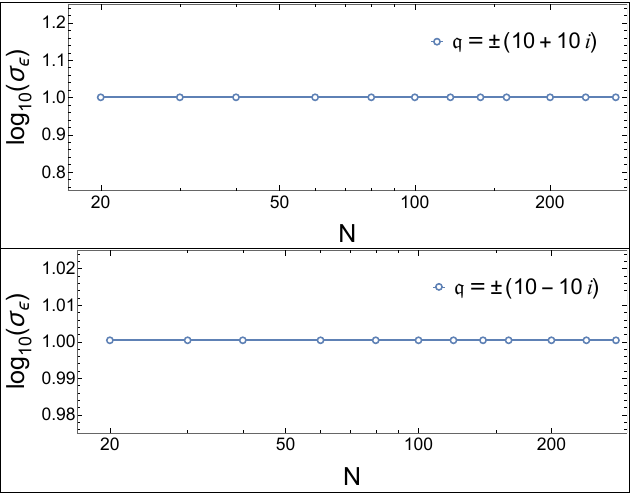}
        \captionsetup{justification=centering}
        \caption{$m^2l^2=-3$, $\mathfrak{w}=0$.}
        \label{fig:Convergencemn3w0}
    \end{subfigure}\hfill
    \begin{subfigure}[b]{0.49\linewidth}
        \centering
        \includegraphics[width=.935\linewidth]{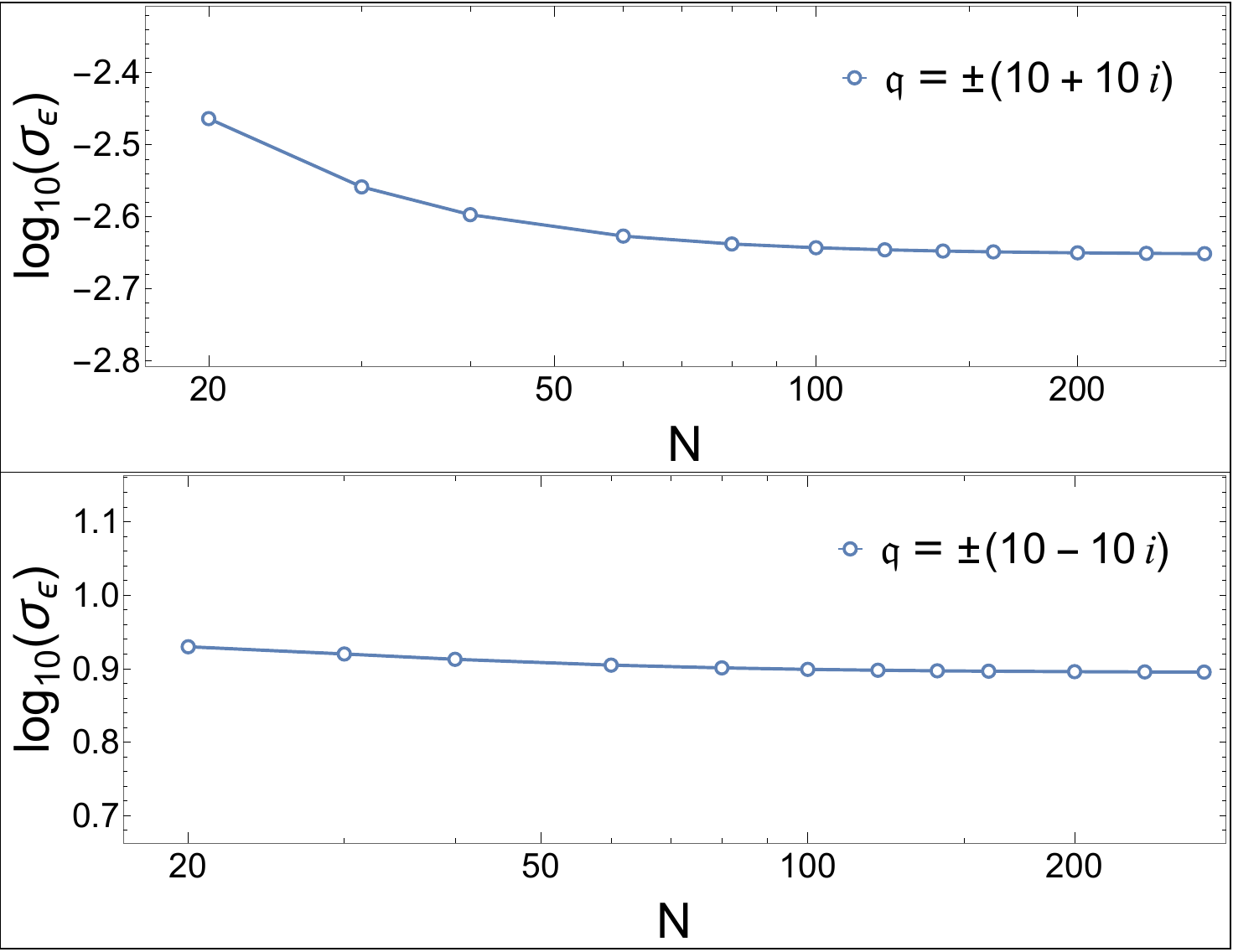}
        \captionsetup{justification=centering}
        \caption{$m^2l^2=-3$, $\mathfrak{w}=10$.}
        \label{fig:Convergencemn3w10}
    \end{subfigure}
    \caption{Norm of the inverse of the resolvent ($\sigma_\epsilon$) as a function of the grid size $N$ evaluated on $\mathfrak{q}=\pm(10\pm10i)$ for different values of $\mathfrak{w}$ and $m^2l^2$. Remarkably, the pseudospectrum converges rapidly and for $N=100$ we can claim that our results should be quantitatively correct up to corrections of order $1\%$.}
    \label{fig:ConvergencePlot}
\end{figure}

\begin{figure}[h]
    \centering
    \begin{subfigure}[b]{0.49\linewidth}
        \centering
        \includegraphics[width=\linewidth]{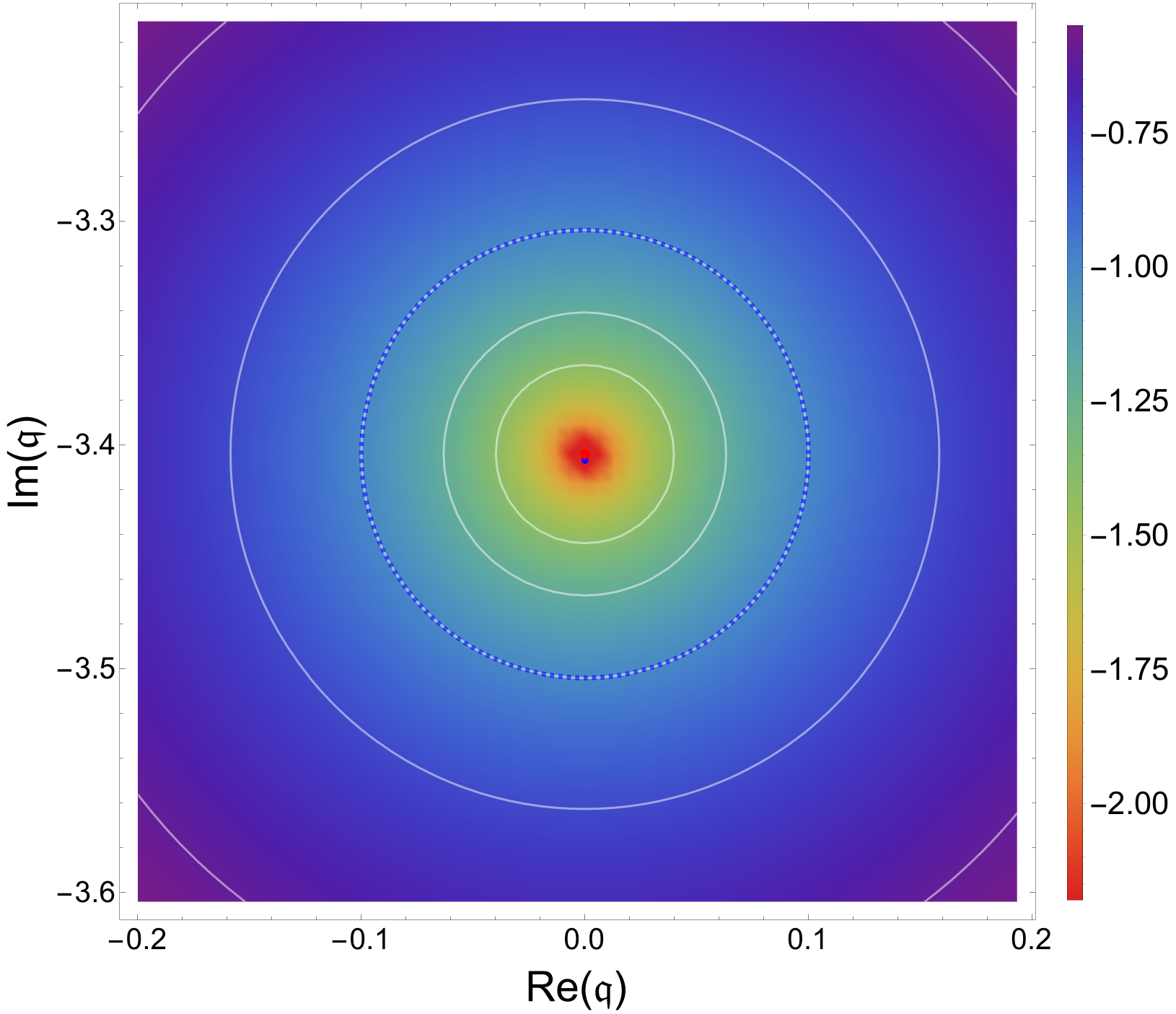}
        \captionsetup{justification=centering}
        \caption{$m^2l^2=0$, $\mathfrak{w}=0$.}
        \label{fig:ScalarAdS5_CloseupPseudo_m0w0}
    \end{subfigure}\hfill
    \begin{subfigure}[b]{0.49\linewidth}
        \centering
        \includegraphics[width=.95\linewidth]{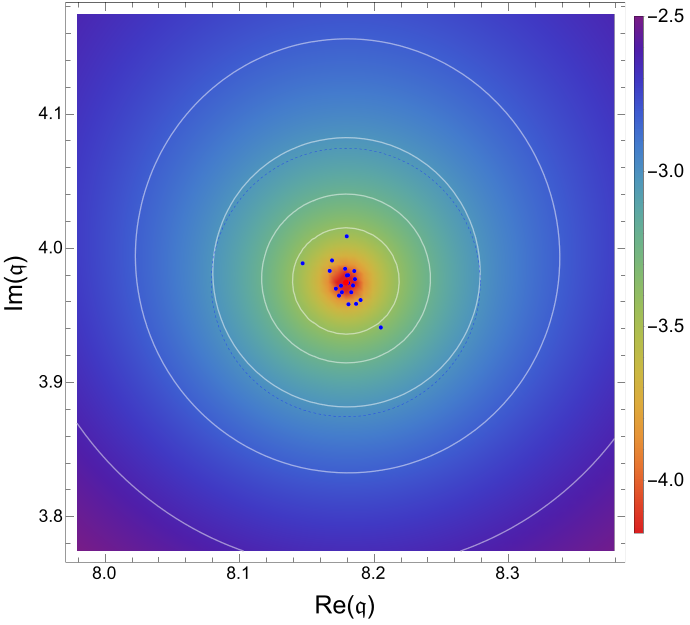}
        \captionsetup{justification=centering}
        \caption{$m^2l^2=0$, $\mathfrak{w}=10$.}
        \label{fig:ScalarAdS5_CloseupPseudo_m0w10}
    \end{subfigure}
    \begin{subfigure}[b]{0.49\linewidth}
        \centering
        \includegraphics[width=\linewidth]{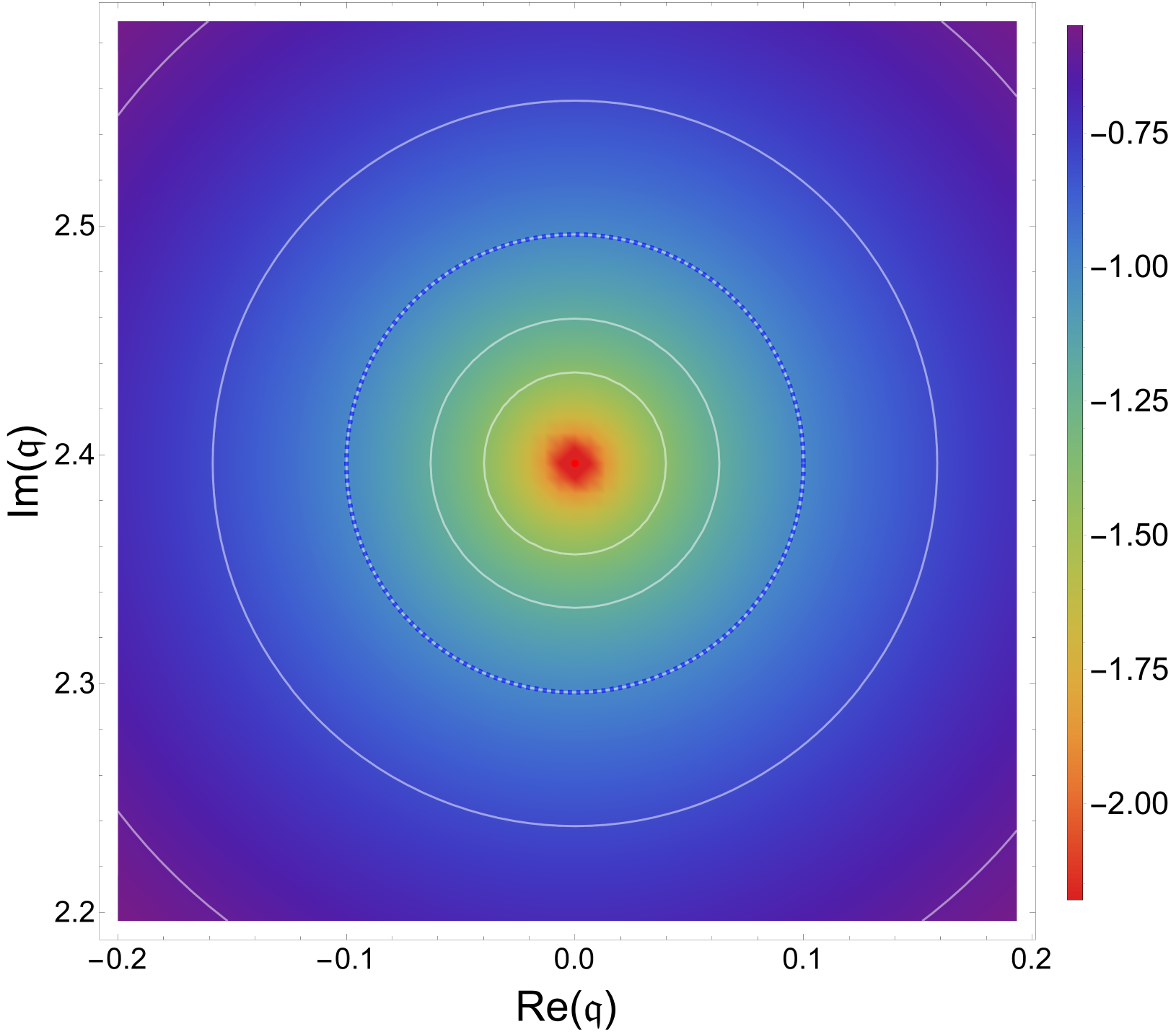}
        \captionsetup{justification=centering}
        \caption{$m^2l^2=-3$, $\mathfrak{w}=0$.}
        \label{fig:ScalarAdS5_CloseupPseudo_mn3w0}
    \end{subfigure}\hfill
    \begin{subfigure}[b]{0.49\linewidth}
        \centering
        \includegraphics[width=\linewidth]{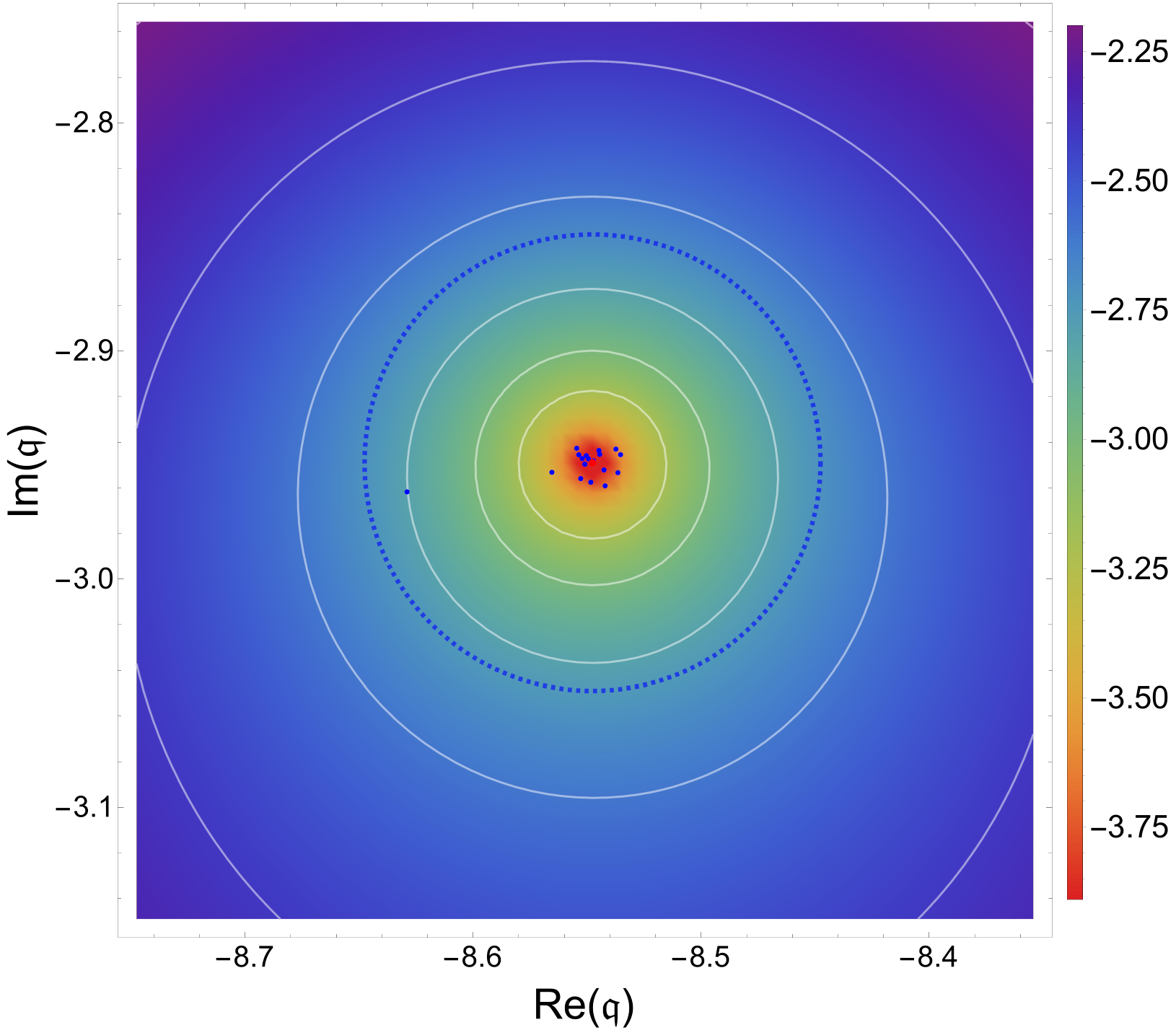}
        \captionsetup{justification=centering}
        \caption{$m^2l^2=-3$, $\mathfrak{w}=10$.}
        \label{fig:ScalarAdS5_CloseupPseudo_mn3w10}
    \end{subfigure}
    \caption{Close-up of the scalar pseudospectrum in the energy norm around the first \CLM for different values of $\mathfrak{w}$ and $m^2l^2$. The red dot corresponds to the \CLM, the white lines represent the boundaries of various full $\varepsilon$-pseudospectra, and the dashed blue circle symbolizes a circle with a radius of $10^{-1}$ centered on the \CLM. The heat map corresponds to the logarithm in base 10 of the inverse of the the norm of the resolvent, while the blue dots indicate selective $\varepsilon$-pseudospectra computed with random local potential perturbations of size $10^{-1}$. Remarkably, in (a) and (c) we observe stability as expected from the compactified theory.}
    \label{fig:CloseupPseudospectraScalar}
\end{figure}

\begin{figure}[htb!]
    \centering
    \begin{subfigure}[b]{0.46\linewidth}
        \centering
        \includegraphics[width=\linewidth]{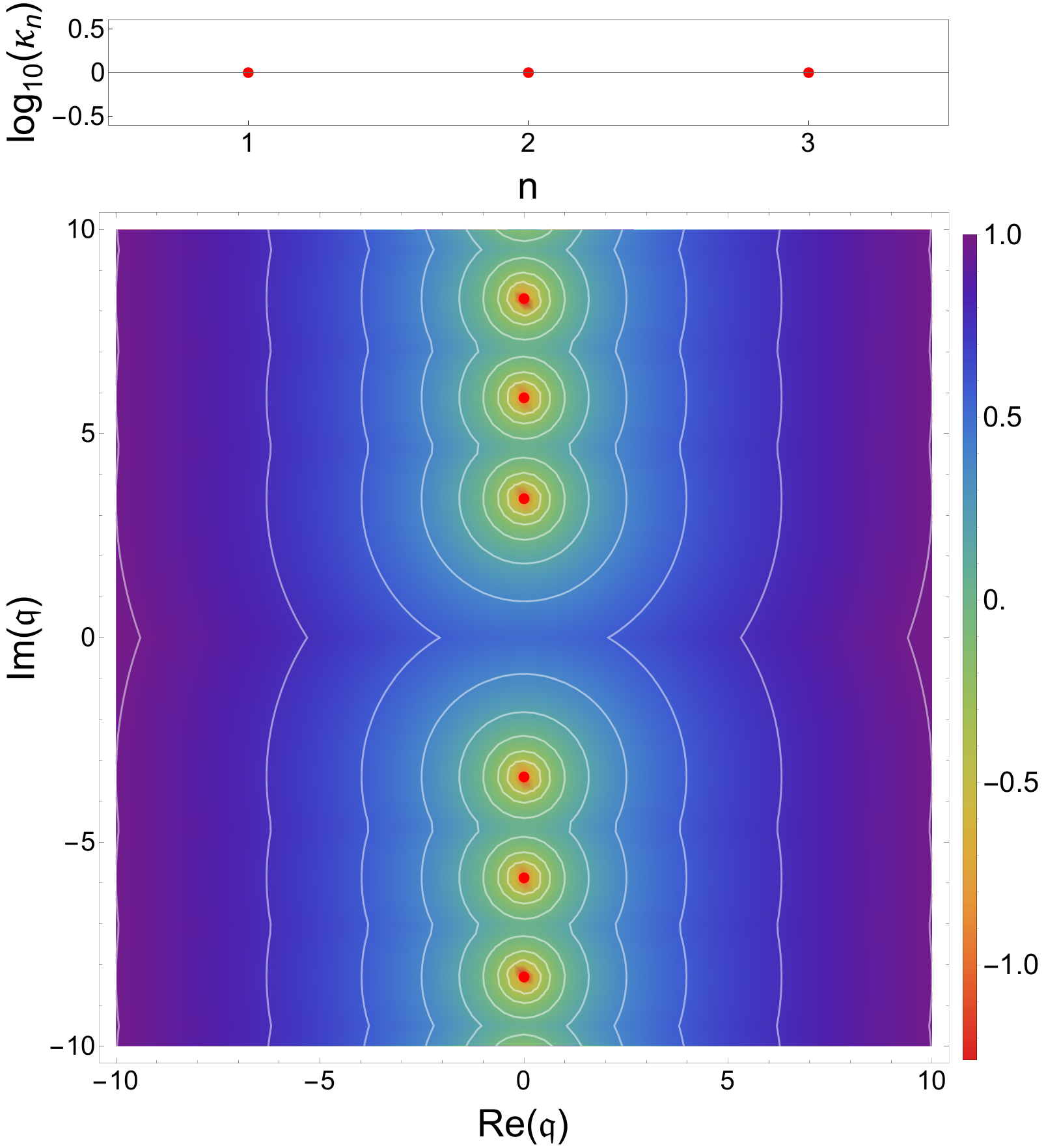}
        \captionsetup{justification=centering}
        \caption{$m^2l^2=0$, $\mathfrak{w}=0$.}
        \label{fig:ScalarAdS5_Pseudo_m0w0}
    \end{subfigure}\hfill
    \begin{subfigure}[b]{0.46\linewidth}
        \centering
        \includegraphics[width=\linewidth]{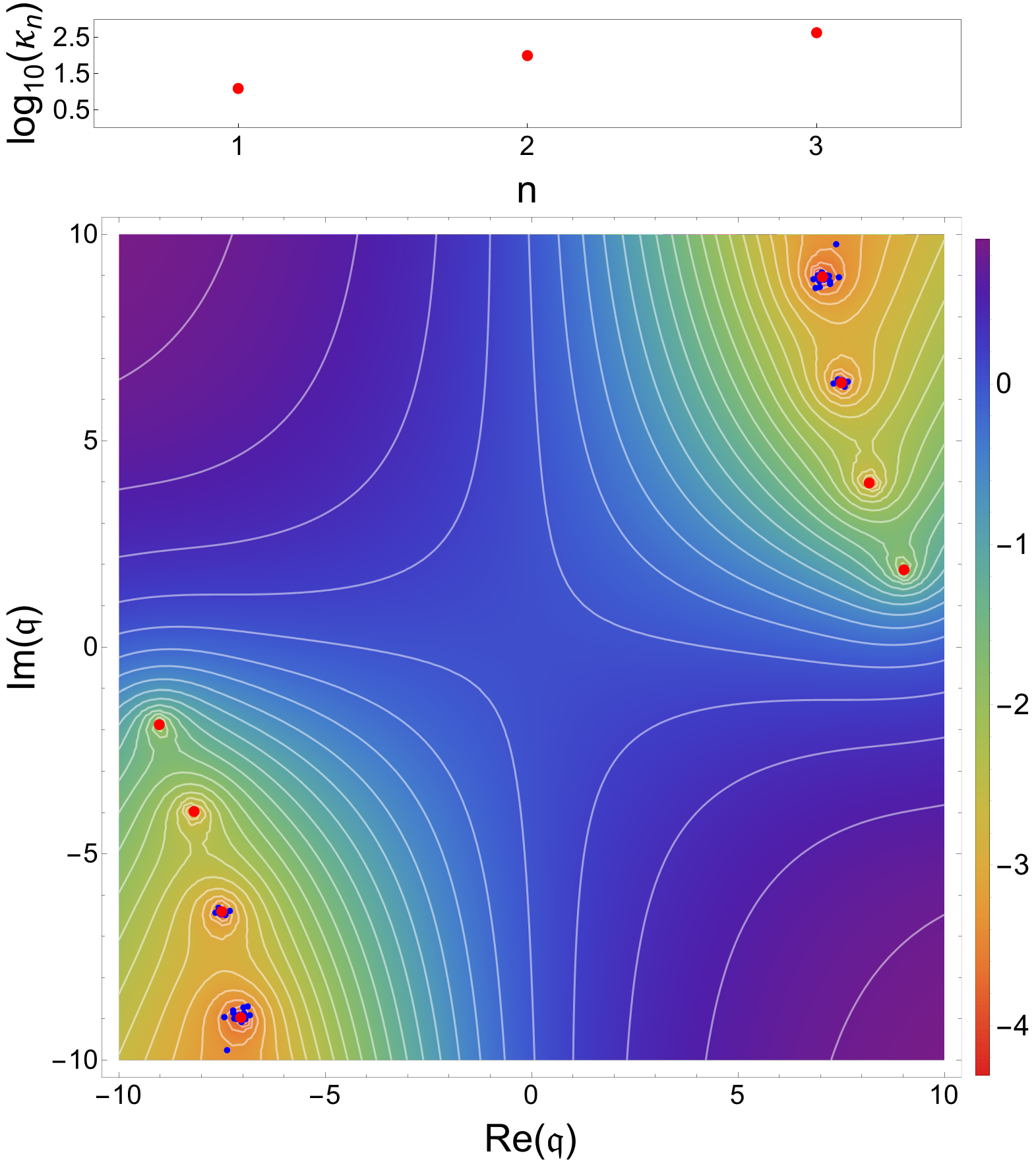}
        \captionsetup{justification=centering}
        \caption{$m^2l^2=0$, $\mathfrak{w}=10$.}
        \label{fig:ScalarAdS5_Pseudo_m0w10}
    \end{subfigure}
    \begin{subfigure}[b]{0.46\linewidth}
        \centering
        \includegraphics[width=\linewidth]{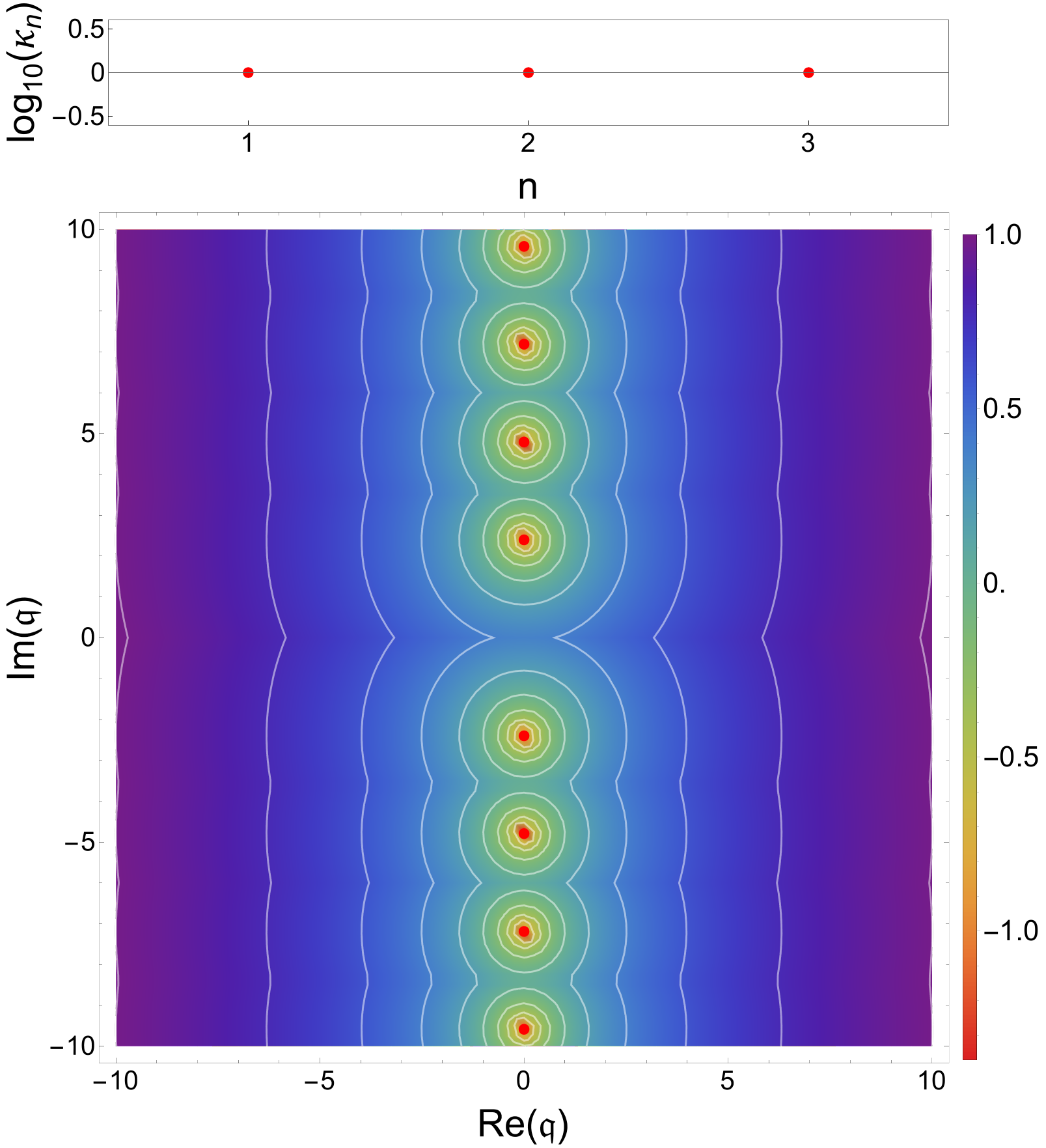}
        \captionsetup{justification=centering}
        \caption{$m^2l^2=-3$, $\mathfrak{w}=0$.}
        \label{fig:ScalarAdS5_Pseudo_mn3w0}
    \end{subfigure}\hfill
    \begin{subfigure}[b]{0.46\linewidth}
        \centering
        \includegraphics[width=\linewidth]{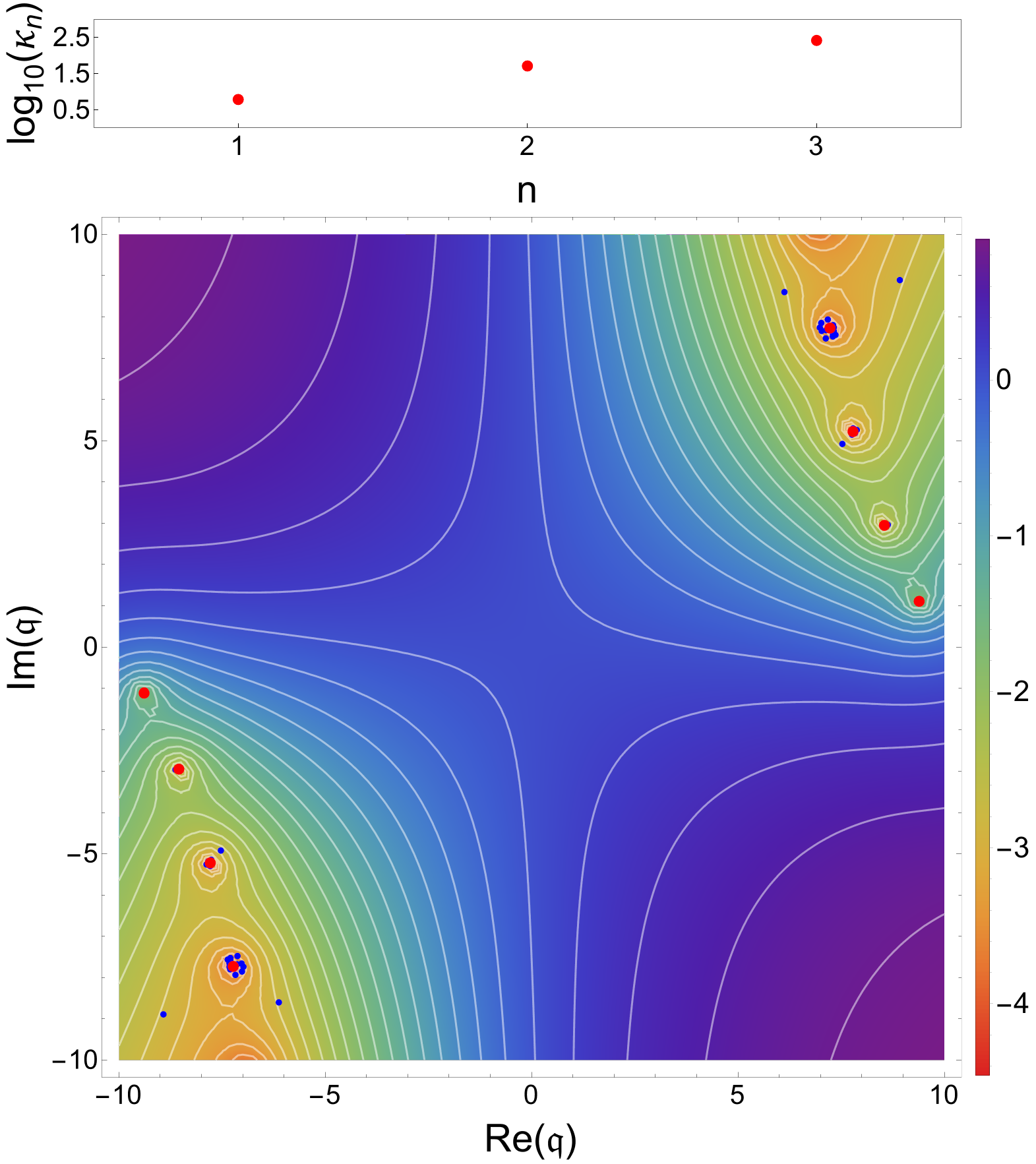}
        \captionsetup{justification=centering}
        \caption{$m^2l^2=-3$, $\mathfrak{w}=10$.}
        \label{fig:ScalarAdS5_Pseudo_mn3w10}
    \end{subfigure}
    \caption{Scalar  pseudospectrum in the energy norm for different values of $\mathfrak{w}$ and $m^2l^2$. In the lower panels, we present selective and full pseudospectra. The red dots represent the (unperturbed) \CLMs.
    The white lines denote the boundaries of different full $\varepsilon$-pseudospectra. The heat map corresponds to the logarithm in base 10 of the inverse of the norm of the resolvent, while the blue dots indicate different selective $\varepsilon$-pseudospectra computed with random local potential perturbations of size $10^{-1}$. In the upper panels, we represent the condition numbers. Most notably, at $\mathfrak{w}\neq0$, for small values of $\varepsilon$, the full $\varepsilon$-pseudospectra present open regions containing multiple QNFs, which signals spectral instability.}
    \label{fig:LargePseudospectraScalar}
\end{figure}

\begin{figure}[htb!]
    \centering
    \begin{subfigure}[b]{0.46\linewidth}
        \centering
        \includegraphics[width=\linewidth]{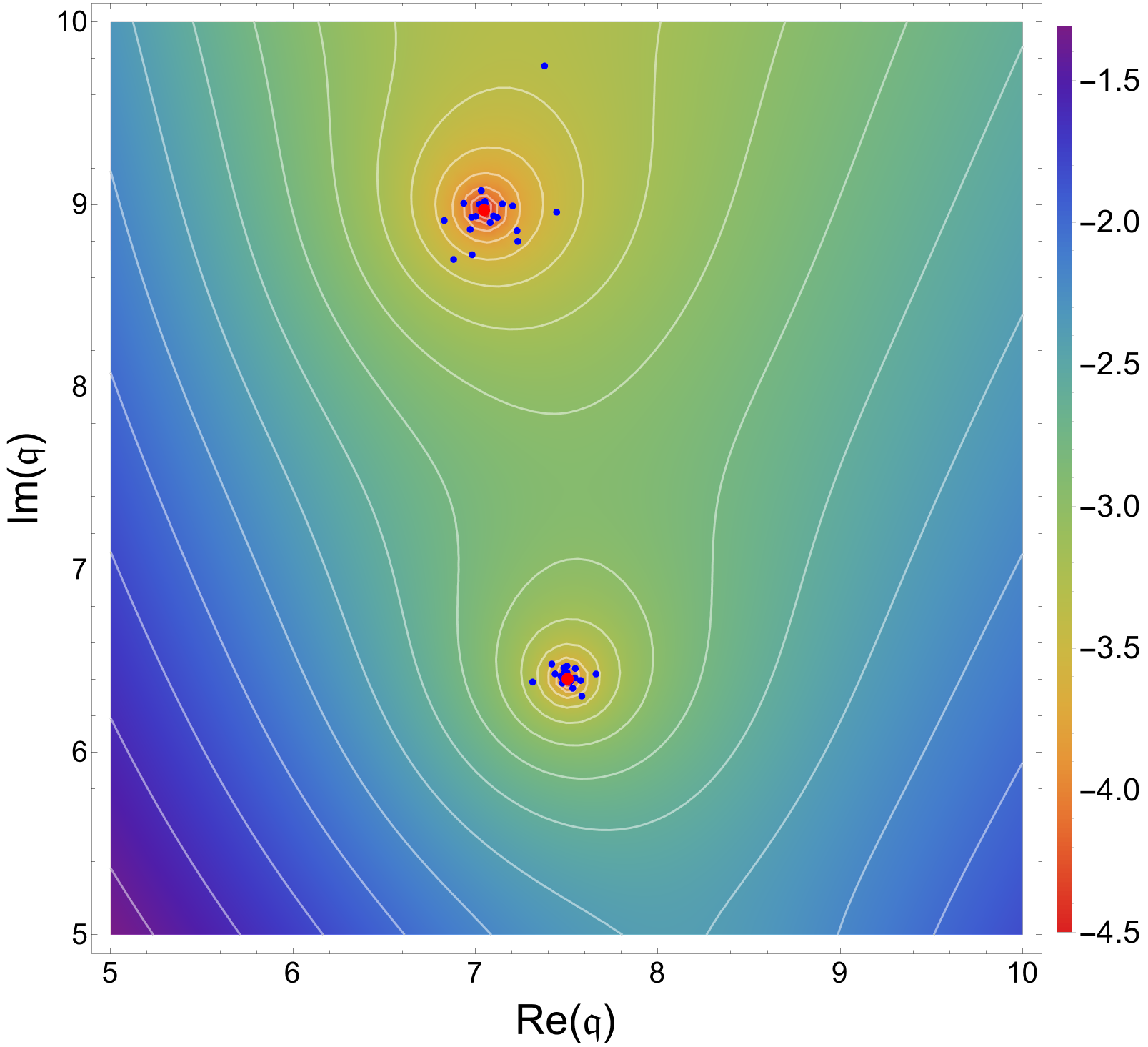}
        \captionsetup{justification=centering}
        \caption{$m^2l^2=0$, $\mathfrak{w}=10$.}
        \label{fig:ScalarAdS5_Pseudo1stQ_m0w10}
    \end{subfigure}\hfill
    \begin{subfigure}[b]{0.46\linewidth}
        \centering
        \includegraphics[width=\linewidth]{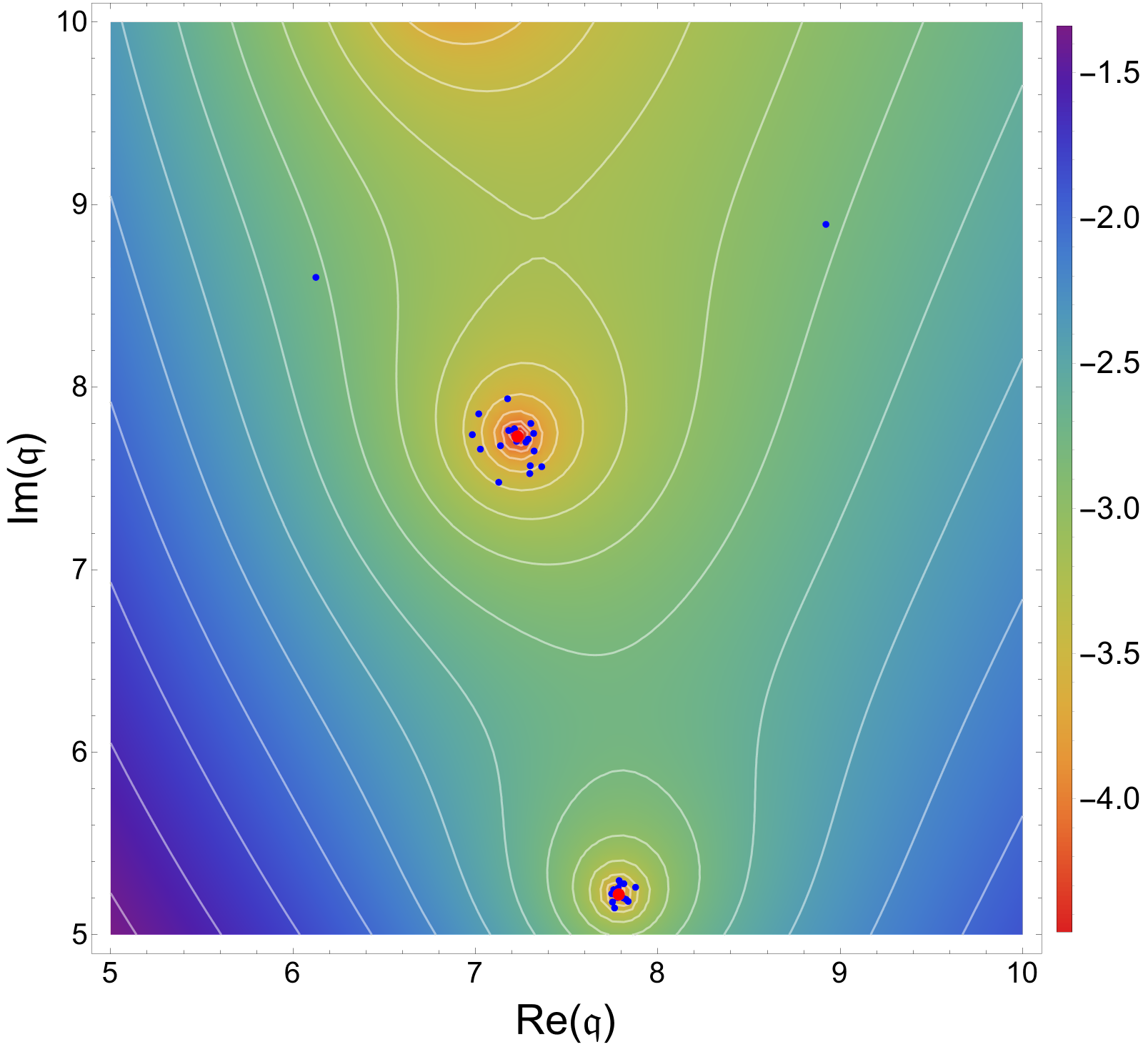}
        \captionsetup{justification=centering}
        \caption{$m^2l^2=-3$, $\mathfrak{w}=10$.}
        \label{fig:ScalarAdS5_Pseudo1stQ_mn3w10}
    \end{subfigure}
    \caption{Close-up of the scalar pseudospectrum in the energy norm around the third and fourth \CLMs for $\mathfrak{w}=10$ and different values of $m^2l^2$. The red dots represent the (unperturbed) \CLMs.
    The white lines denote the boundaries of different full $\varepsilon$-pseudospectra. The heat map corresponds to the logarithm in base 10 of the inverse of the norm of the resolvent, while the blue dots indicate different selective $\varepsilon$-pseudospectra computed with random local potential perturbations of size $10^{-1}$. Remarkably the instability under random local potential perturbations probed by the selective pseudospectrum is orders of magnitude milder than that of generic perturbations probed by the full pseudospectrum.}
    \label{fig:LargePseudospectraScalar1stQ}
\end{figure}

\begin{figure}[h!]
    \centering
    \begin{subfigure}[b]{0.49\linewidth}
        \centering
        \includegraphics[width=\linewidth]{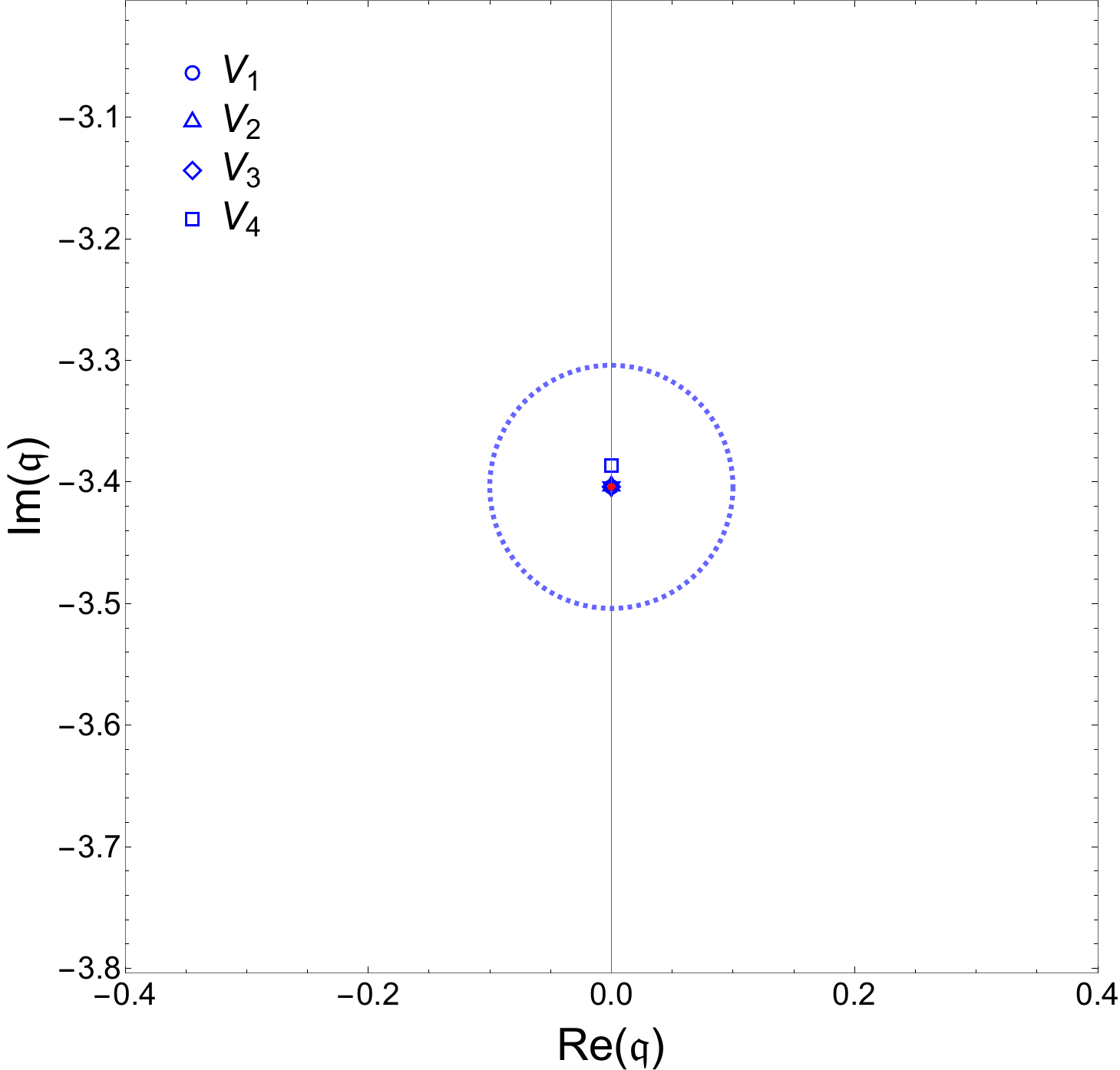}
        \captionsetup{justification=centering}
        \caption{$m^2l^2=0$, $\mathfrak{w}=0$.}
        \label{fig:ScalarAdS5_Deterministic_m0w0}
    \end{subfigure}\hfill
    \begin{subfigure}[b]{0.49\linewidth}
        \centering
        \includegraphics[width=.98\linewidth]{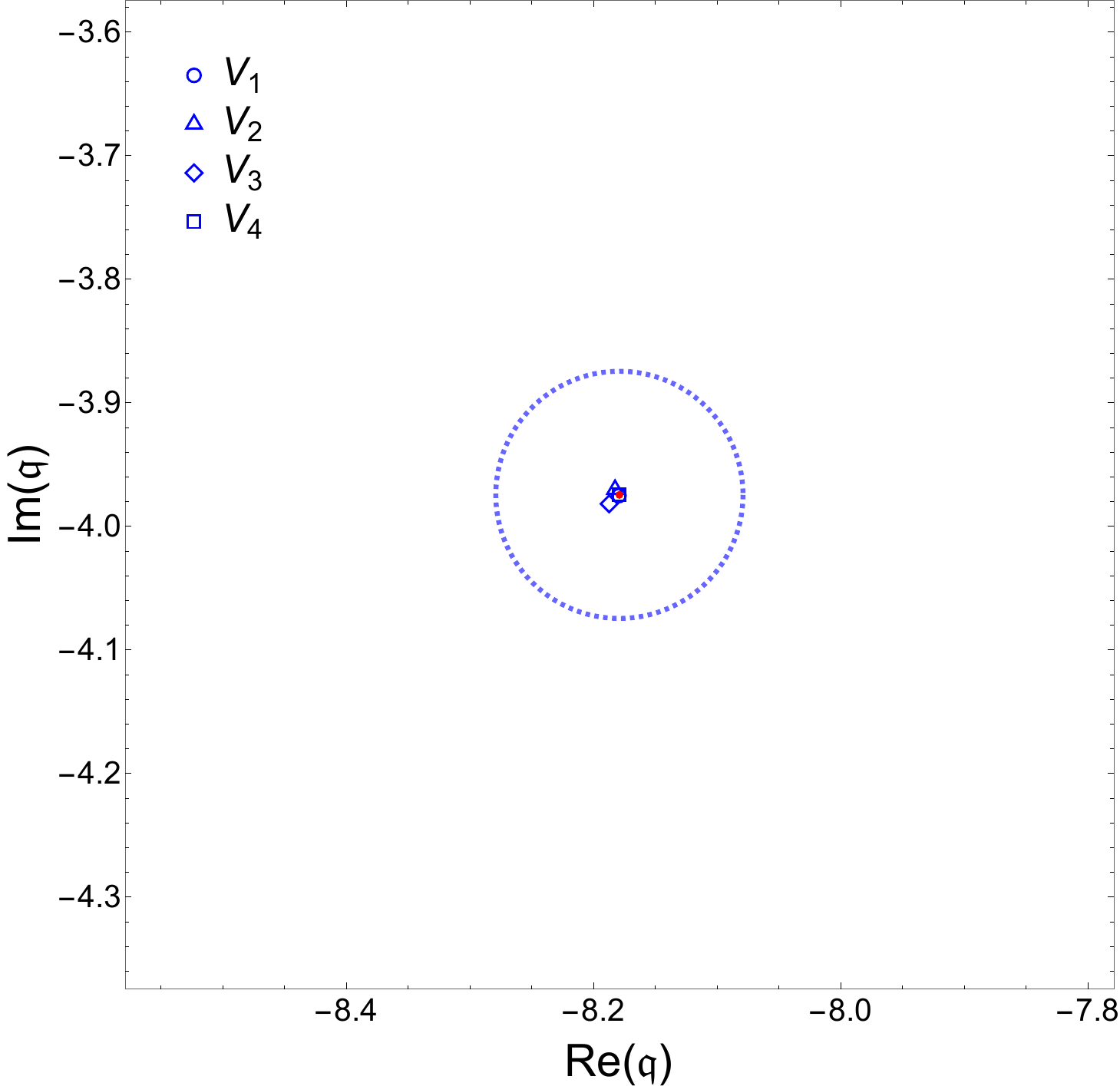}
        \captionsetup{justification=centering}
        \caption{$m^2l^2=0$, $\mathfrak{w}=10$.}
        \label{fig:ScalarAdS5_Deterministic_m0w10}
    \end{subfigure}
    \begin{subfigure}[b]{0.49\linewidth}
        \centering
        \includegraphics[width=\linewidth]{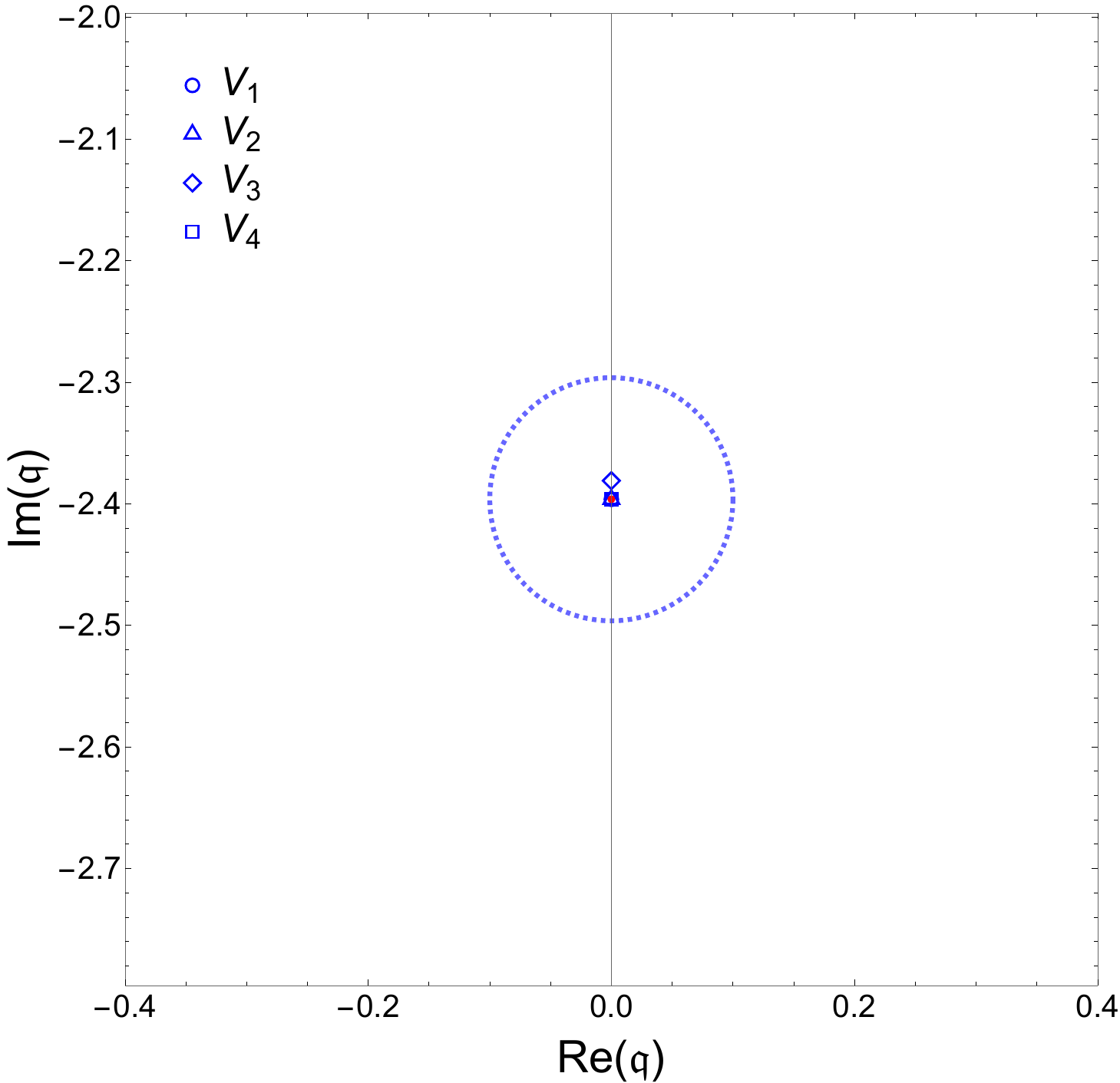}
        \captionsetup{justification=centering}
        \caption{$m^2l^2=-3$, $\mathfrak{w}=0$.}
        \label{fig:ScalarAdS5_Deterministic_mn3w0}
    \end{subfigure}\hfill
    \begin{subfigure}[b]{0.49\linewidth}
        \centering
        \includegraphics[width=.98\linewidth]{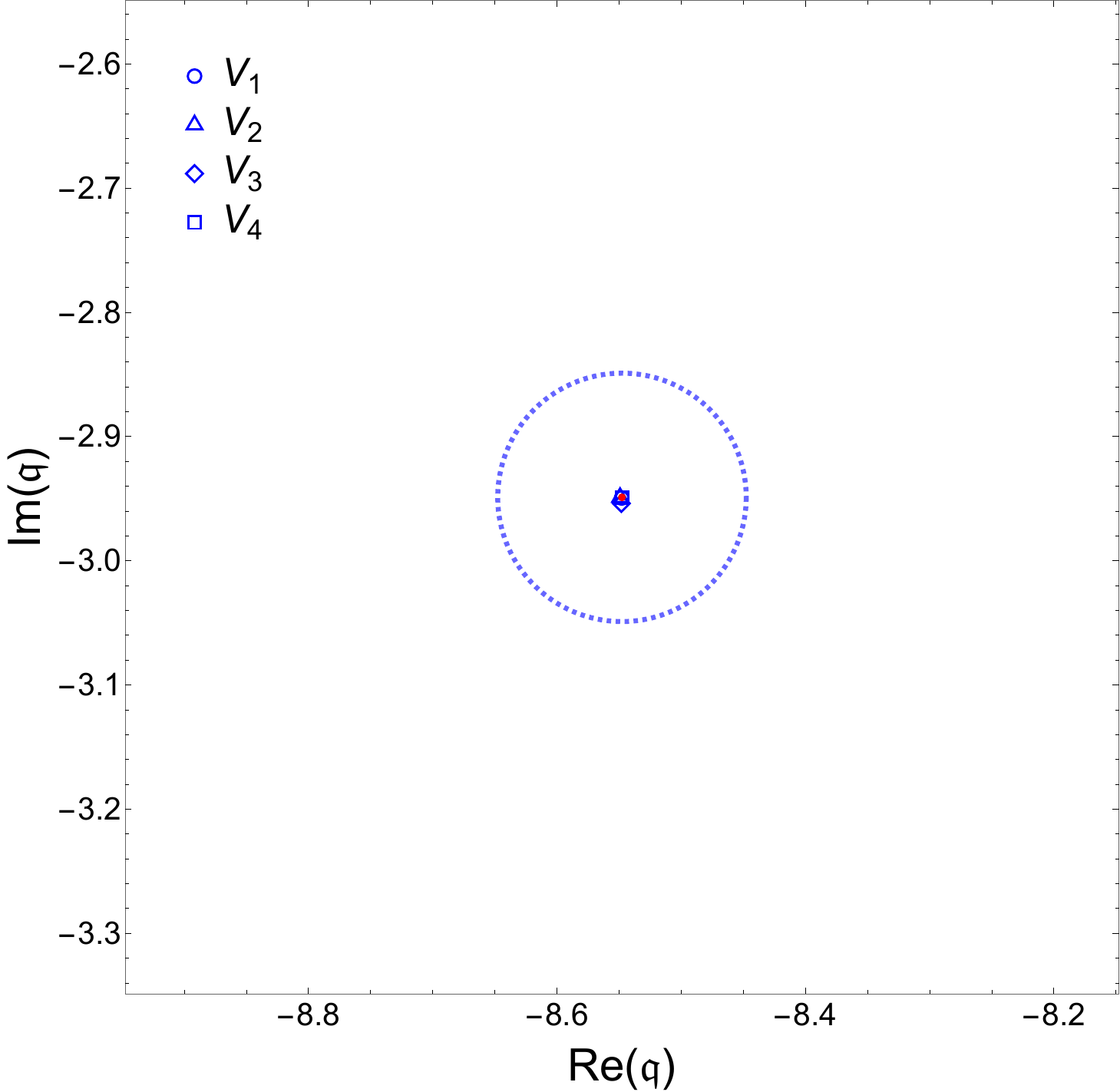}
        \captionsetup{justification=centering}
        \caption{$m^2l^2=-3$, $\mathfrak{w}=10$.}
        \label{fig:ScalarAdS5_Deterministic_mn3w10}
    \end{subfigure}
    \caption{Effect on \CLMs of the deterministic perturbations \eqref{eq:deterministicpots} with size $\norm{V_i}_E=10^{-1}$. The unperturbed \CLM is shown in red, while the perturbed \CLMs are depicted in blue. The dashed blue line represents the circle of radius $10^{-1}$ centered in the unperturbed \CLM. Remarkably, even for $\mathfrak{w}\neq0$ we observe stability under deterministic perturbations.}
    \label{fig:CloseupDeterministicScalar}
\end{figure}   

\begin{figure}[h!]
    \centering
    \begin{subfigure}[b]{0.44\linewidth}
        \centering
        \includegraphics[width=\linewidth]{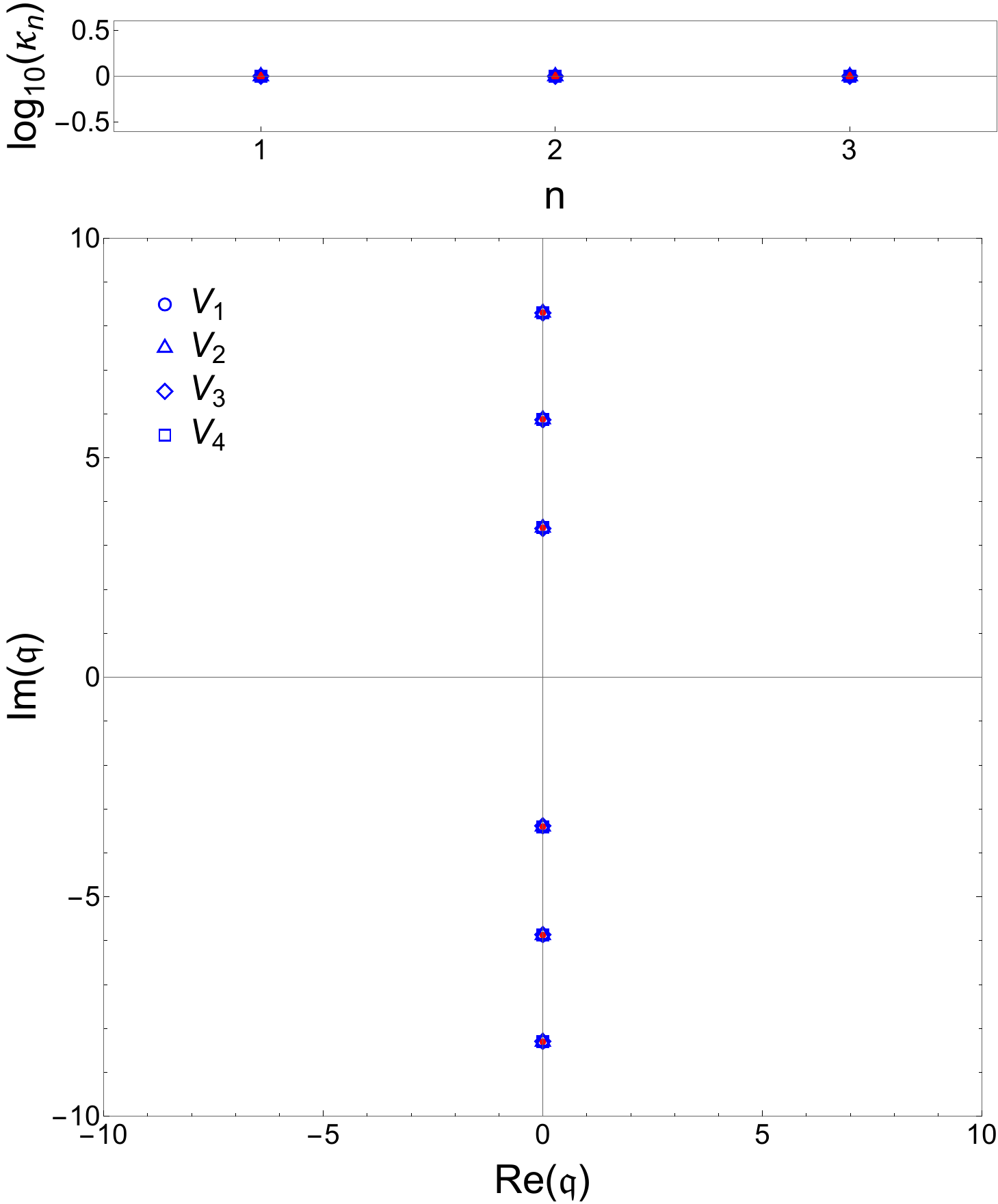}
        \captionsetup{justification=centering}
        \caption{$m^2l^2=0$, $\mathfrak{w}=0$.}
        \label{fig:ScalarAdS5_LargeDeterministic_m0w0}
    \end{subfigure}\hfill
    \begin{subfigure}[b]{0.44\linewidth}
        \centering
        \includegraphics[width=\linewidth]{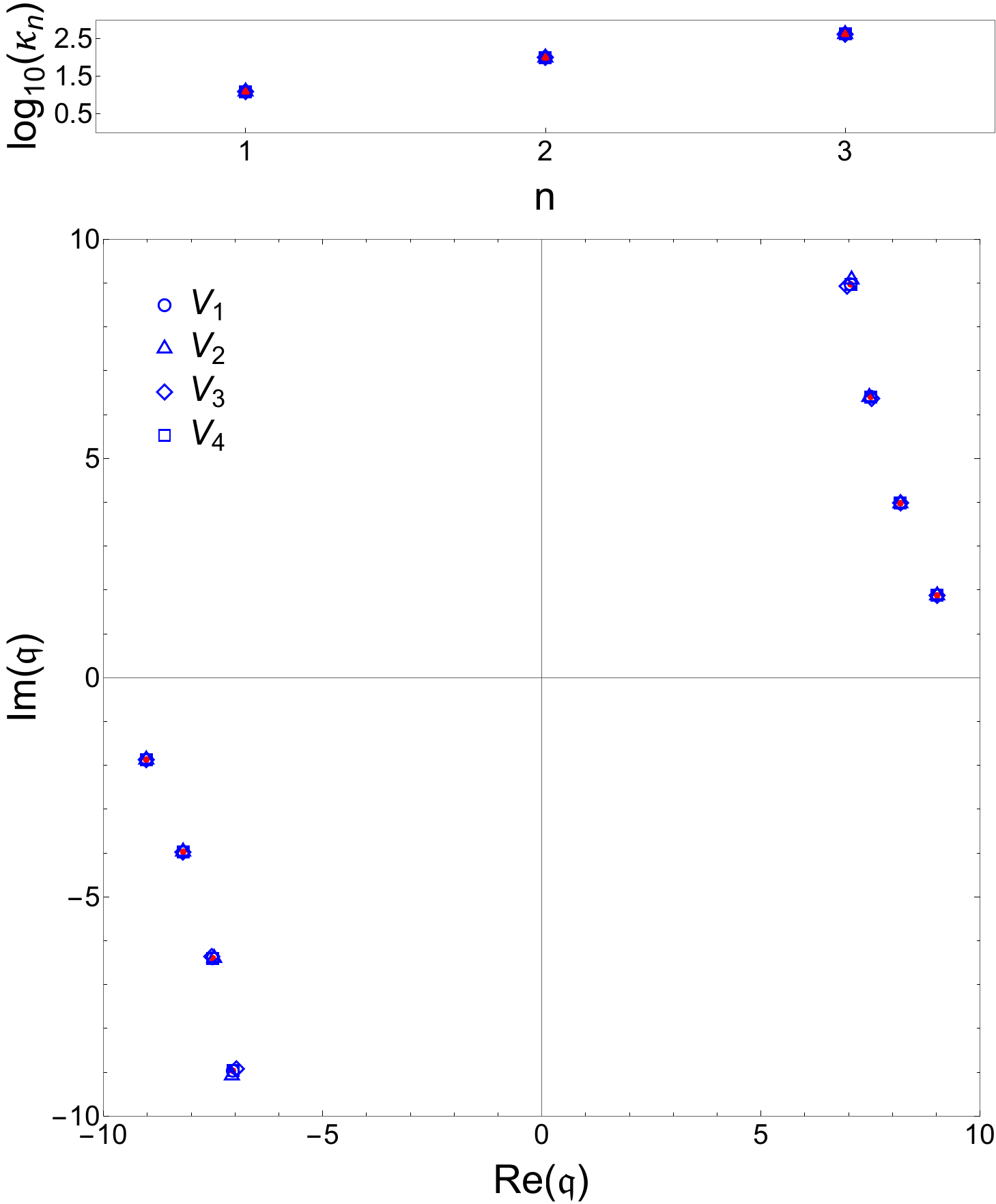}
        \captionsetup{justification=centering}
        \caption{$m^2l^2=0$, $\mathfrak{w}=10$.}
        \label{fig:ScalarAdS5_LargeDeterministic_m0w10}
    \end{subfigure}
    \begin{subfigure}[b]{0.44\linewidth}
        \centering
        \includegraphics[width=\linewidth]{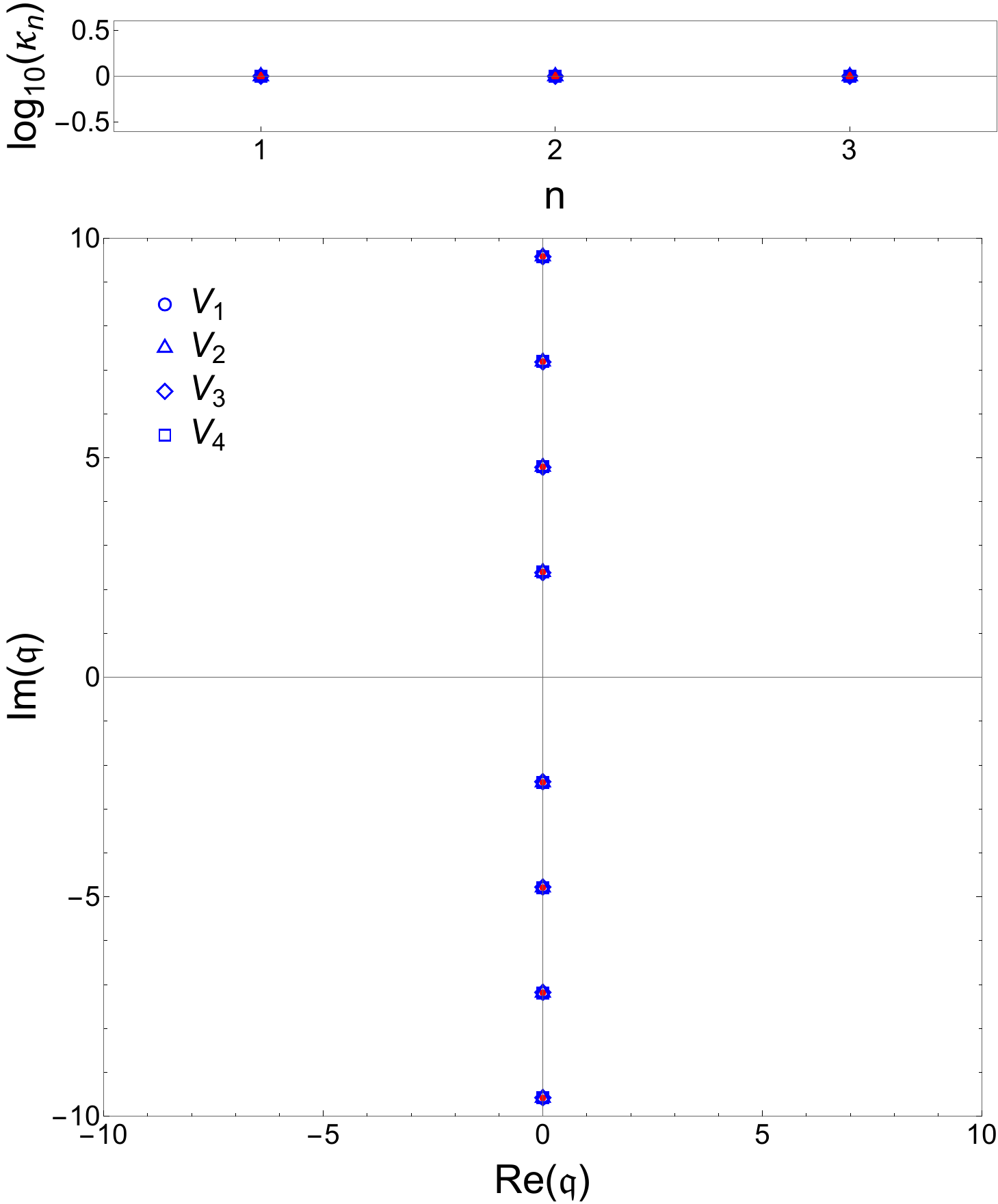}
        \captionsetup{justification=centering}
        \caption{$m^2l^2=-3$, $\mathfrak{w}=0$.}
        \label{fig:ScalarAdS5_LargeDeterministic_mn3w0}
    \end{subfigure}\hfill
    \begin{subfigure}[b]{0.44\linewidth}
        \centering
        \includegraphics[width=\linewidth]{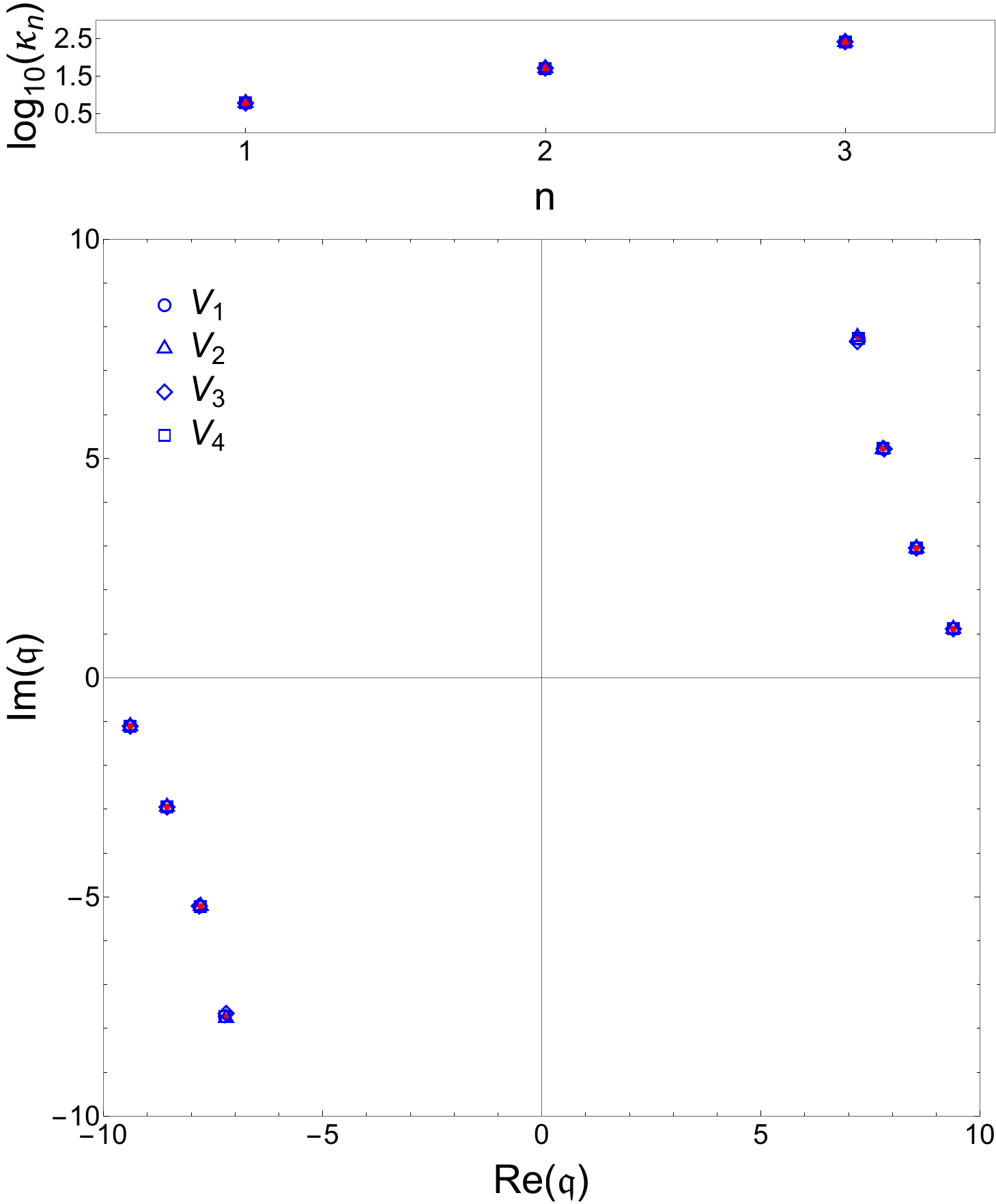}
        \captionsetup{justification=centering}
        \caption{$m^2l^2=-3$, $\mathfrak{w}=10$.}
        \label{fig:ScalarAdS5_LargeDeterministic_mn3w10}
    \end{subfigure}
    \caption{Effect on the spectrum of the scalar of the deterministic perturbations %
    \eqref{eq:deterministicpots}
    with size $\norm{V_i}_E=10^{-1}$. In the lower panels we present the spectra and in the upper ones the condition numbers for the lowest \CLMs. The unperturbed \CLMs are shown in red, while the perturbed ones are depicted in blue. In the plotted region of the spectrum is stable under all perturbations even for $\mathfrak{w}\neq0$.}
    \label{fig:LargeDeterministicScalar}
\end{figure}

\begin{figure}[htb!]
    \centering
    \begin{subfigure}[b]{0.49\linewidth}
    \centering
    \includegraphics[width=\linewidth]{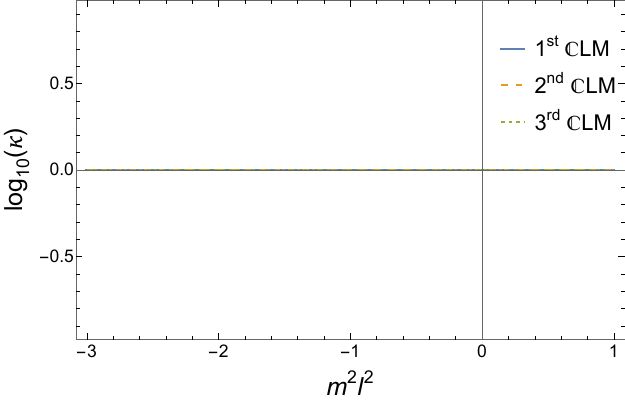}
    \captionsetup{justification=centering}
    \caption{$\mathfrak{w}=0$.}
    \label{fig:ConditionNumbersw0}
    \end{subfigure}\hfill
    \begin{subfigure}[b]{0.49\linewidth}
    \centering
    \includegraphics[width=\linewidth]{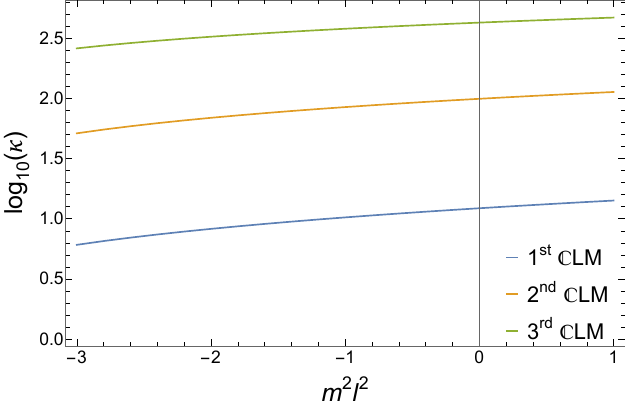}
    \captionsetup{justification=centering}
    \caption{$\mathfrak{w}=10$.}
    \label{fig:ConditionNumbersw10}
    \end{subfigure}
    \caption{Mass dependence of the condition numbers of the complex momenta. Note that for $\mathfrak{w}=0$ all condition numbers are one.}
    \label{fig:mDependence}
\end{figure}

\begin{figure}[htb!]
    \centering
    \begin{subfigure}[b]{0.49\linewidth}
    \centering
    \includegraphics[width=\linewidth]{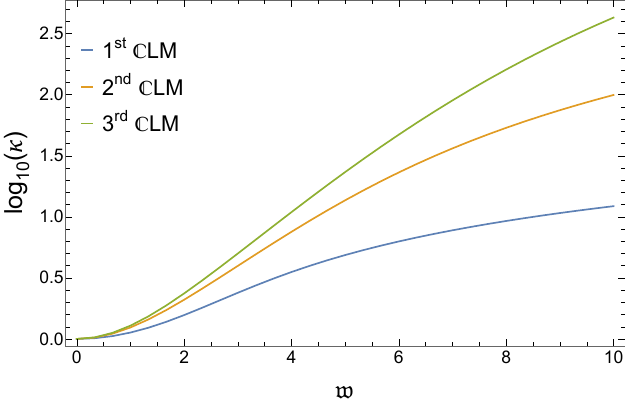}
    \captionsetup{justification=centering}
    \caption{$m^2l^2=0$.}
    \label{fig:ConditionNumbersm0}
    \end{subfigure}\hfill
    \begin{subfigure}[b]{0.49\linewidth}
    \centering
    \includegraphics[width=\linewidth]{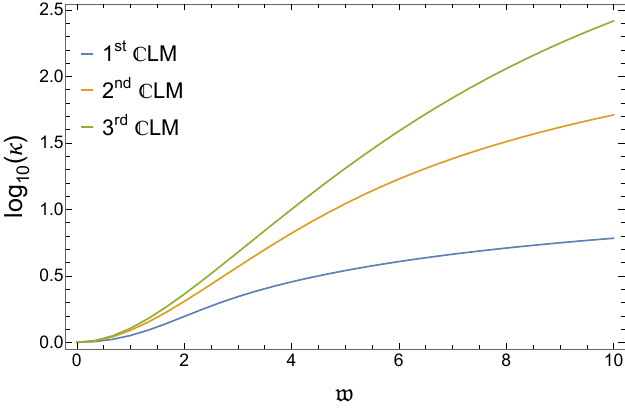}
    \captionsetup{justification=centering}
    \caption{$m^2l^2=-3$.}
    \label{fig:ConditionNumbersmn3}
    \end{subfigure}
    \caption{Frequency dependence of the condition numbers of the \CLMs.}
    \label{fig:wDependence}
\end{figure}

\clearpage

\section{Comparison with quasinormal frequencies at complex momentum}\label{app:Comparisons}

In this appendix we explicitly show the differences in the stability properties of QNFs and \CLMs. As we argued in the main text, while related, QNFs and \CLMs appear in very different physical settings. Thus, the stability properties of a mode depend heavily on whether we consider it a QNM or a \CMM.

\begin{figure}[htb!]
    \centering
    \begin{subfigure}[b]{0.49\linewidth}
    \centering
    \includegraphics[width=\linewidth]{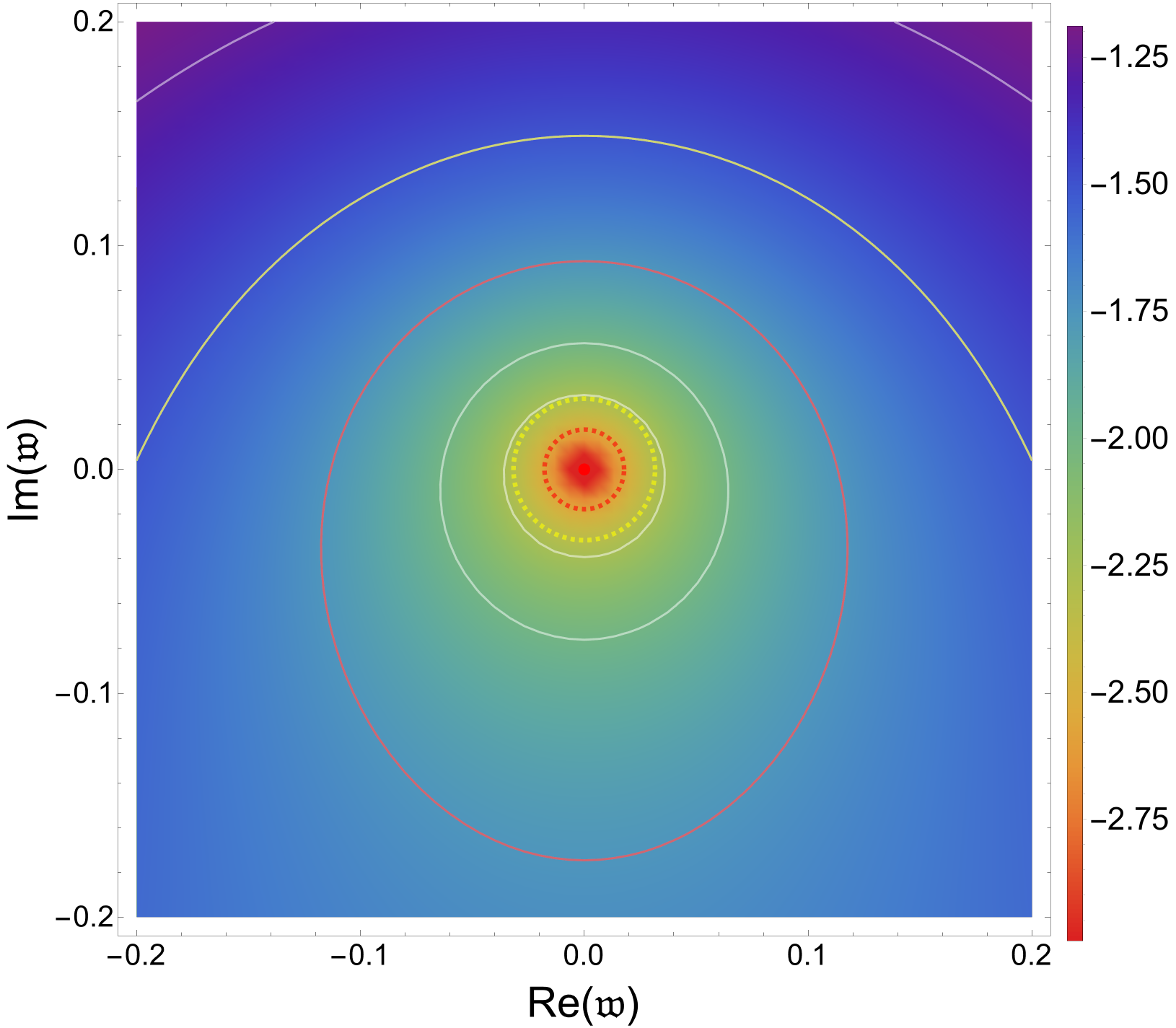}
    \captionsetup{justification=centering}
    \caption{QNF $\mathfrak{w}=0$ at $\mathfrak{q}=-3.40407 i$.}
    \label{fig:QNFvsCLM_QNF}
    \end{subfigure}\hfill
    \begin{subfigure}[b]{0.49\linewidth}
    \centering
    \includegraphics[width=\linewidth]{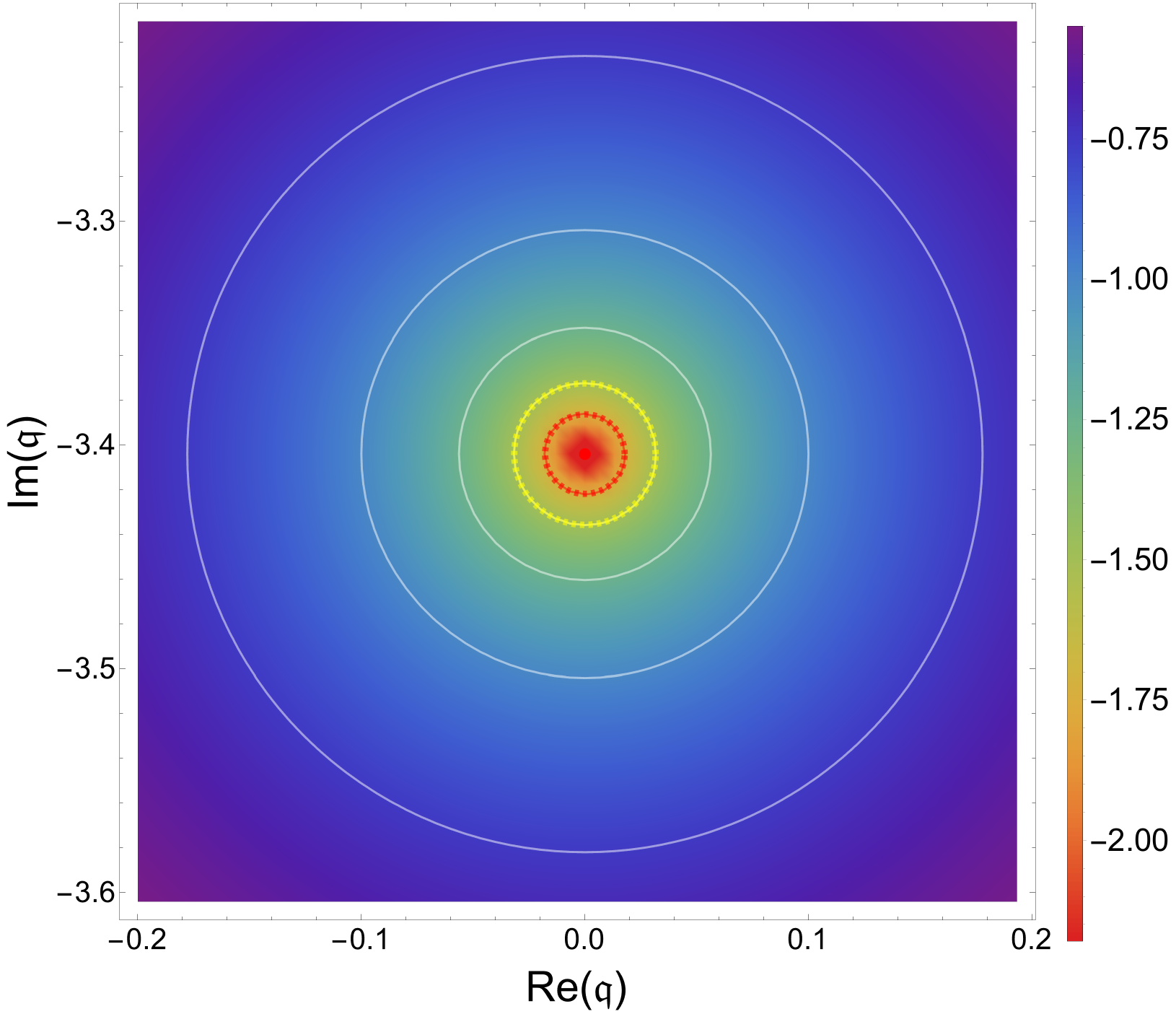}
    \captionsetup{justification=centering}
    \caption{\CLM $\mathfrak{q}=-3.40407 i$ at $\mathfrak{w}=0$.}
    \label{fig:QNFvsCLM_CLM}
    \end{subfigure}
    \caption{Comparison of the pseudospectra of a QNF and a \CLM. The red dots represent the eigenvalues. The heat map corresponds to the logarithm in base 10 of the inverse of the norm of the resolvent. The white lines denote the boundaries $\varepsilon$-pseudospectra and the red (yellow) lines correspond to the boundaries of the $10^{-3/2}$ ($10^{-7/4}$) pseudospectra. The dashed red (yellow) circle represents the circle of radius $10^{-3/2}$ ($10^{-7/4}$) centered around the eigenvalue. Note that in \ref{fig:QNFvsCLM_CLM} the spectrum is stable as the dashed red (yellow) circle coincides with the boundary of the $10^{-3/2}$ ($10^{-7/4}$) pseudospectra. On the other hand, in \ref{fig:QNFvsCLM_QNF} the spectrum is unstable as the boundary of the $10^{-3/2}$ ($10^{-7/4}$) pseudospectra is much larger than the dashed red (yellow) circle.}
    \label{fig:QNFvsCLM}
\end{figure}

To explicitly show this we consider the scalar field \eqref{eq:Action scalar} with $m=0$. This scalar has a \CLM $\mathfrak{q}=-3.40407 i$ at $\mathfrak{w}=0$ and, correspondingly, a QNF $\mathfrak{w}=0$ at $\mathfrak{q}=-3.40407 i$. To analyze the differences in stability we compare the pseudospectra of the \CLM and of the QNF in figure \ref{fig:QNFvsCLM}. Clearly we can see that the \CLM is stable, while the QNF is not. Thus showcasing their different stability properties.

The \CLM pseudospectrum is computed following the procedure discussed throughout this paper. The differential operator is given by \eqref{eq:Def L for doublet} and the inner product by \eqref{eq:InnerProduct}. On the other hand, to compute the QNF pseudospectrum we consider the following eigenvalue problem and norm introduced in \cite{Arean:2023ejh}
\begin{equation}
    \mathfrak{w}\Psi = i
    \begin{pmatrix}
    0&1 \\
    L_1\left[\partial_\rho^2,\partial_\rho;\boldsymbol{\mathfrak{q}},\rho\right]&L_2\left[\partial_\rho;\boldsymbol{\mathfrak{q}},\rho\right]   
    \end{pmatrix}\Psi\,,
\end{equation}
\begin{equation}
    \expval{\Psi_1,\Psi_2}_E=\int \frac{d\rho}{(1-\rho)^3}\,\, \left[ 
        \mathfrak{|q|}^2\bar{\phi}_1\phi_2+f\partial_\rho \bar{\phi}_1\partial_\rho\phi_2+(2-f)\bar{\psi_1}\psi_2\right]\,,
\end{equation}
where $\Psi=(\phi,z_h\partial_t\phi)=(\phi,\psi)$ and the differential operators $L_1$ and $L_2$ take the form:
\begin{align}\label{eq:Eigenvalue problem real scalar 2}
    L_1\left[\partial_\rho^2,\partial_\rho;\boldsymbol{\mathfrak{q}},\rho\right]&=\left[f-2\right]^{-1}\left[\boldsymbol{\mathfrak{q}}^2-(1-\rho)^3\left(\frac{f }{(1-\rho)^3}\right)'\partial_{\rho} -f\partial_{\rho}^2 \right]\,,\\[\medskipamount]\label{eq:Eigenvalue problem real scalar 3}
    L_2\left[\partial_\rho;\boldsymbol{\mathfrak{q}},\rho\right]&=\left[f-2\right]^{-1}\left[(1-\rho)^3\left(\frac{f-1}{(1-\rho)^3}\right)' +2 \left(f-1\right)\partial_{\rho}  \right]\,.
\end{align}

\printbibliography

\end{document}